\documentclass[namedreferences]{solarphysics}

\usepackage[hyperref,optionalrh]{spr-sola-addons} 
\usepackage{graphicx}        
\usepackage{amssymb}        
\usepackage{natbib}
\usepackage{booktabs}
\usepackage{multirow}
\usepackage[usenames,dvipsnames]{color}           
\usepackage{breakurl}        

\chardef\us=`\_

\begin{document}

\begin{article}
\begin{opening}

\title{Computation of Relative Magnetic Helicity in Spherical Coordinates}

\author[addressref=aff1,email={kostas.moraitis@obspm.fr}]{\inits{K.}\fnm{Kostas}~\lnm{Moraitis}}
\author[addressref=aff1]{\inits{E.}\fnm{\'Etienne}~\lnm{Pariat}}
\author[addressref=aff2]{\inits{A.}\fnm{Antonia}~\lnm{Savcheva}}
\author[addressref=aff3]{\inits{G.}\fnm{Gherardo}~\lnm{Valori}}

\address[id=aff1]{LESIA, Observatoire de Paris, PSL Research University, CNRS, Sorbonne Universit\'{e}s, UPMC Univ. Paris 06, Univ. Paris Diderot, Sorbonne Paris Cit\'{e}, 5 place Jules Janssen, 92195 Meudon, France}
\address[id=aff2]{Harvard-Smithsonian Center for Astrophysics, 60 Garden Street, Cambridge, MA 02138, USA}
\address[id=aff3]{Mullard Space Science Laboratory, University College London, 
   		Holmbury St. Mary, Dorking, Surrey, RH5 6NT, UK}

\runningauthor{K. Moraitis \textit{et al.}}
\runningtitle{Computation of Relative Magnetic Helicity in Spherical Coordinates}

\begin{abstract}
Magnetic helicity is a quantity of great importance in solar studies because it is conserved in ideal magneto-hydrodynamics. While many methods to compute magnetic helicity in Cartesian finite volumes exist, in spherical coordinates, the natural coordinate system for solar applications, helicity is only treated approximately. We present here a method to properly compute relative magnetic helicity in spherical geometry. The volumes considered are finite, of shell or wedge shape, and the three-dimensional magnetic field is considered fully known throughout the studied domain. Testing of the method with well-known, semi-analytic, force-free magnetic-field models reveals that it has excellent accuracy. Further application to a set of nonlinear force-free reconstructions of the magnetic field of solar active regions, and comparison with an approximate method used in the past, indicates that the proposed methodology can be significantly more accurate, thus making our method a promising tool in helicity studies that employ the spherical geometry. Additionally, the range of applicability of the approximate method is determined and discussed.
\end{abstract}


\end{opening}

\section{Introduction}
     \label{S-Introduction} 

Magnetic helicity is a geometrical quantity that describes the twist, writhe, and linkage of magnetic-field lines. It is invariant in ideal magneto-hydrodynamics \citep[MHD;][]{woltjer58}, and is approximatelly conserved even in nonideal MHD conditions \citep{taylor74,pariat15}. These properties make helicity an important quantity in plasma physics studies \citep[\textit{e.g.}][]{ji95,canfield99,dasgupta02}. 

Magnetic helicity of a magnetic field [$\mathbfit{B}$] in a volume [$V$] is defined as $H_m(\mathbfit{B})=\int_V\,\mathrm{d}V \mathbfit{A} \cdot \mathbfit{B}$, where $\mathbfit{A}$ is the vector potential such that $\mathbfit{B}=\nabla\times\mathbfit{A}$. This quantity is gauge-invariant as long as the surface enclosing $V$ is a flux surface, \textit{i.e.} when the normal component of the magnetic field vanishes there. This is obviously not the case for arbitrarily shaped volumes, but still an appropriate helicity with respect to a reference field can be defined \citep{BergerF84,fa85}. This is the relative magnetic helicity, which is given by
\begin{equation}
H=\int_V\,\mathrm{d}V (\mathbfit{A}+\mathbfit{A}_\mathrm{p})\cdot (\mathbfit{B}-\mathbfit{B}_\mathrm{p}),
\label{helc}
\end{equation}
with $\mathbfit{B}_\mathrm{p}=\nabla\times\mathbfit{A}_\mathrm{p}$ the reference field, usually taken to be a potential field, and $\mathbfit{A}_\mathrm{p}$ its vector potential. The important condition allowing gauge independence is that the studied field and the potential field must have the same normal component on the \emph{whole} boundary of the system.

The range of applications of relative magnetic helicity (hereafter simply helicity) in the Sun is quite broad, extending from the solar dynamo \citep[\textit{e.g.}][]{brandenburg05} to the triggering of coronal mass ejections: according to \citet{rust94}, the existence of solar eruptions might be attributed to the need to shed the coronal magnetic helicity, constantly accumulating from a continuous injection through the photosphere.

In the solar context, different methods and approaches can be used to evaluate magnetic helicity. A review and comparison of several of these methods is presented by \citet{valori16}. The focus of this article is on the finite volume methods \citep[\textit{e.g.}][]{thalmann11,val12,yang13}, that is, methods that employ the definition of helicity as a volume integral and thus require the three-dimensional (3D) magnetic field in the entire volume as input. 

In all of these methods, the computation of relative magnetic helicity is performed in Cartesian geometry. The natural coordinate system for the Sun however is the spherical one, and so it is necessary to be able to calculate helicity in spherical coordinates. Possible applications of such a calculation include MHD simulations in spherical geometry \citep[\textit{e.g.}][]{masson13,fan16,karpen17}, as well as nonlinear force-free (NLFF) field extrapolations \citep[\textit{e.g.}][]{Gilchrist2014,savcheva16}.

Although there are studies that compute magnetic helicity in spherical coordinates \citep[\textit{e.g.}][]{bobra08,fan10,savcheva15,karpen17}, these use various assumptions and/or simplifications. The respective helicity computations are then problem-dependent, and the methods used cannot be generalised to other datasets. In the MHD simulations of \citet{fan10}, for example, there is no magnetic flux penetrating the lateral boundaries of the volume, which results in the simplification of the helicity calculations. In another example, \citet{karpen17} drive a coronal hole jet with a rotational photospheric plasma motion that leaves approximately unchanged the initially prescribed potential magnetic field (and corresponding vector potential), thus simplifying the helicity calculations.

There are other cases, as in \citet{bobra08}, where a different helicity formula is used. In spherical coordinates this reads
\begin{equation}
H_R=\int_V\,\mathrm{d}V (\mathbfit{A}\cdot \mathbfit{B} - \mathbfit{A}_\mathrm{p}\cdot \mathbfit{B}_\mathrm{p})+\int_S\,\mathrm{d}S \chi B_r.
\label{helb}
\end{equation}
This relation stems from the original definition of relative helicity \citep{BergerF84} where space is divided in two regions; the volume of interest [$V$], and a complementary volume [$V'$] where $\mathbfit{B}'=\mathbfit{B}_\mathrm{p}'$, and thus $\mathbfit{A}'-\mathbfit{A}'_\mathrm{p}=\nabla \chi$ for some scalar function $\chi$ (with primed quantities referring to $V'$). The interface of the two volumes is denoted as $S$, while $B_r$ is the radial component of the field. Additionally, the boundary of $V\cup V'$ is considered as a flux surface, or extending to infinity.

In the particular case examined by \citet{bobra08}, $V$ is the coronal wedge-shaped volume of interest, and $V'$ its sub-photospheric extension to the center of the Sun, and so neither of them is in general bounded by a flux surface. Following however \citet{bobra08}, $V'$ is a bounded volume, and since $V\cup V'$ is bounded by a flux surface, it follows that $V'$ should be bounded by a flux surface below the photosphere. This necessarily implies that the photospheric boundary $S$ has to be flux-balanced, \textit{i.e.} without net magnetic flux. These restrictions could be lifted by considering $V'$ as the whole volume outside $V$, but this was not the choice made in the original derivation of Equation~\ref{helb}, as the limitation of the surface integral to the photosphere demonstrates. Indeed, the first term in Equation~\ref{helb} represents the contribution to helicity from $V$, while the same quantity in $V'$ reduces to the surface integral of the second term because of the assumption $\mathbfit{B}'=\mathbfit{B}_\mathrm{p}'$ there. Had a different geometry of $V'$ been adopted, an additional contribution to Equation~\ref{helb} should be considered.

To see whether, and under which conditions, Equation~\ref{helc} is equivalent to Equation~\ref{helb}, we apply the latter to the same volume as the former, $V\cup V'$. The contribution to relative magnetic helicity from $V'$ is however identicaly zero, since $\mathbfit{B}'=\mathbfit{B}_\mathrm{p}'$ there. In the volume of interest, $V$, we can expand the terms in Equation~\ref{helc} as
\begin{equation}
H=\int_V\,\mathrm{d}V (\mathbfit{A}\cdot \mathbfit{B} - \mathbfit{A}_\mathrm{p}\cdot \mathbfit{B}_\mathrm{p})+\int_V\,\mathrm{d}V\,(\mathbfit{A}_\mathrm{p}\cdot \mathbfit{B} - \mathbfit{A}\cdot \mathbfit{B}_\mathrm{p}).
\label{helc0}
\end{equation}
The first term coincides with the respective term in Equation~\ref{helb}, while the second is a mixed term which can also be written as the surface integral
\begin{equation}
H_{\rm mix}=\oint_{\partial V}\,\mathrm{d}S\, \hat{\mathbfit{n}}\cdot (\mathbfit{A}\times\mathbfit{A_{\rm p}}),
\label{helc1}
\end{equation}
after using standard vector idendities \citep{val12}. Here $\hat{\mathbfit{n}}$ stands for the outward-pointing unit vector on the boundary of the volume of interest $\partial V$.

To better compare the surface terms of the two helicity equations, we reverse the steps in the derivation of the second term of Equation~\ref{helb}, and get
\begin{equation}
\int_S\mathrm{d}S\, \chi B_r=\int_{\partial V'}\mathrm{d}S\, \chi  (\hat{\mathbfit{n}}'\cdot\mathbfit{B}')=\int_{V'}\mathrm{d}V\,\nabla\chi\cdot\mathbfit{B}'.
\end{equation}
In the first equality the surface integral is extended to the whole boundary of the sub-photospheric volume, $\partial V'$, by the assumption that $\partial V'-S$ is a flux surface, $\hat{\mathbfit{n}}'\cdot\mathbfit{B}'=0$, with $\hat{\mathbfit{n}}'$ denoting the outward-pointing unit vector on $\partial V'$. The second equality follows directly from Gauss' theorem and an integration by parts. Now, since $\nabla\chi\cdot\mathbfit{B}'=(\mathbfit{A}'-\mathbfit{A}'_\mathrm{p})\cdot\mathbfit{B}'=-\nabla\cdot(\mathbfit{A}'\times\mathbfit{A'_{\rm p}})$ in $V'$, we derive
\begin{equation}
\int_S\mathrm{d}S \chi B_r=-\int_{\partial V'}\mathrm{d}S\, \hat{\mathbfit{n}}'\cdot(\mathbfit{A}'\times\mathbfit{A'_{\rm p}})=-\int_S\mathrm{d}S\, \hat{\mathbfit{n}}'\cdot(\mathbfit{A}'\times\mathbfit{A'_{\rm p}}).
\label{helbt1}
\end{equation}
Again, only the photospheric part survives in the surface integral since the remaining boundaries are flux surfaces. Moreover, as noted by \citet{fa85}, Equation~\ref{helb} is gauge-invariant if additionally the tangential components of the two vector potentials are continuous across $S$, a condition that leads to $\mathbfit{A}'\times\mathbfit{A'_{\rm p}}=\mathbfit{A}\times\mathbfit{A_{\rm p}}$ on $S$. Replacing in Equation~\ref{helbt1} also that $\hat{\mathbfit{n}}'=-\hat{\mathbfit{n}}$ on $S$, we finally get
\begin{equation}
\int_S\mathrm{d}S \chi B_r=\int_S\mathrm{d}S\, \hat{\mathbfit{n}}\cdot(\mathbfit{A}\times\mathbfit{A_{\rm p}}).
\label{helbt2}
\end{equation}
The surface term in the \citet{bobra08} formulation thus corresponds to the photospheric part of the $H_{\rm mix}$ term.

The comparison of the surface terms of the two helicity formulations, given by Equations~\ref{helbt2} and \ref{helc1}, shows that they coincide if the condition $\hat{\mathbfit{n}}\cdot (\mathbfit{A}\times\mathbfit{A_{\rm p}}) =0$ holds on the coronal part of the boundary of the volume of interest, $\partial V-S$. The conditions for Equations~\ref{helc} and \ref{helb} to be equivalent are thus that $\left. \hat{\mathbfit{n}}\cdot (\mathbfit{A}\times\mathbfit{A_{\rm p}}) \right|_{\partial V-S}=0$ on the coronal part of $\partial V$, and $\left. \hat{\mathbfit{n}}'\cdot\mathbfit{B}'\right|_{\partial V'-S}=0$ on the sub-photospheric part of $\partial V'$, also implying magnetic flux balance on $S$. The first condition is obviously gauge-dependent, unless $V$ extents to infinity (where the vector potentials presumably vanish), or $\partial V$ is a flux surface (so that the tangential components of $\mathbfit{A}$ vanish there). If additionally the sub-photospheric part of $\partial V'$ is treated as in \citet{bobra08}, \textit{i.e.} it is assumed as generic and not bounded by a flux surface, with $S$ not flux balanced, then the condition $\hat{\mathbfit{n}}\cdot (\mathbfit{A}\times\mathbfit{A_{\rm p}})=0$ must be valid also there, and then $H_{\rm mix}=0$ holds throughout the boundary of $V\cup V'$.

In deriving Equation~\ref{helb} therefore, the gauge-dependent assumption $H_{\rm mix}=0$ is made implicitly. This should be taken into account when applying this equation to finite volumes, since it is a gauge-dependent condition that is not valid in general.

In this article we extend the computationally most efficient and robust of the finite-volume methods in Cartesian coordinates \citep{val12,moraitis14} to the spherical geometry. The equations that we derive are applicable to both spherical-shell and spherical-wedge geometries. However, we focus on the latter because it is of more common use. We also compare the results produced by this method with helicity values computed with an approximate method based on Equation~\ref{helb}, and we determine the applicability of the latter. In Section~\ref{S-helicity} we describe the implementation details of the method and in Section~\ref{S-validation} we perform its validation against semi-analytic force-free magnetic field models. In Section~\ref{S-application} we apply it to a set of NLFF fields that are also described there, and finally in Section~\ref{S-discussion} we summarize and discuss the results of the article.

\section{Relative Magnetic Helicity in Spherical Coordinates}
     \label{S-helicity}

Relative magnetic helicity is computed directly from its definition by \citet{fa85}, Equation~\ref{helc}. From a given 3D magnetic field [$\mathbfit{B}$] one has to estimate the other three vector fields appearing in Equation~\ref{helc}, namely the potential magnetic field [$\mathbfit{B}_\mathrm{p}$], and the corresponding vector potentials [$\mathbfit{A}$ and $\mathbfit{A}_\mathrm{p}$] of the two magnetic fields. Note that helicity as given in Equation~\ref{helc} is independent of the gauges used in the vector potentials as long as the normal components of the original and potential fields match along the \emph{whole} boundary of the volume, \textit{i.e.}
\begin{equation}
\left.\hat{\mathbfit{n}}\cdot\mathbfit{B}\right|_{\partial V}=\left.\hat{\mathbfit{n}}\cdot\mathbfit{B}_\mathrm{p}\right|_{\partial V} .
\label{batbound}
\end{equation}
The calculation of relative magnetic helicity can then be done in two steps, similarly to the Cartesian case \citep{val12,moraitis14}, as detailed in the present section.

\subsection{Calculation of the Potential Field}
     \label{S-helicity-pf} 

The potential field can always be expressed through a scalar potential $\Phi$ as $\mathbfit{B}_\mathrm{p}=\nabla \Phi$. The potential then satisfies Laplace's equation $\nabla^2\Phi=0$ in $V$. In spherical coordinates this reads
\begin{equation}
\frac{1}{r^2} \frac{\partial}{\partial r}  \left( r^2 \frac{\partial \Phi}{\partial r} \right)+\frac{1}{r^2\sin \theta} \frac{\partial}{\partial \theta} \left( \sin\theta \frac{\partial \Phi}{\partial \theta} \right)+\frac{1}{r^2\sin \theta^2} \frac{\partial^2 \Phi}{\partial \phi^2}=0.
\label{lapleq}
\end{equation}
The requirement of Equation~\ref{batbound}, ensuring gauge-invariance, leads to the following Neumann boundary conditions for the scalar potential
\begin{equation}
\left.\frac{\partial\Phi}{\partial\hat{\mathbfit{n}}}\right|_{\partial V}=\left.\hat{\mathbfit{n}}\cdot\mathbfit{B}\right|_{\partial V}.
\label{neumbc}
\end{equation}
We note that the solution of Laplace's equation under Neumann boundary conditions exists (up to an additive constant) only for flux-balanced fields, $\int_{\partial V}\mathbfit{B}\cdot\mathrm{d}\mathbfit{S}=0$, a condition that is never fully satisfied with numerical data. 

Laplace's equation is solved with the use of a \textsf{FORTRAN} routine contained in the \textsf{MUDPACK} library \citep{adams89}. The routine uses a multigrid iteration method to obtain the solution and is therefore efficient, quick, and computationally not demanding. The multigrid method requires a uniform computational grid that has dimensions of the special form $m\, 2^n +1$, for $n$ some positive integer, and $m$ a small prime number, like 2, 3, or 5. In the case that either of these conditions is not fulfilled by the input magnetic field data, we linearly interpolate them to an appropriate grid, and we inevitably introduce numerical errors to the following helicity calculations.

\subsection{Calculation of the Vector Potentials}
     \label{S-helicity-vp} 

\subsubsection{Analytical Derivation}

For the calculation of the vector potential of a given 3D magnetic field we follow the conceptual method initially developed by \citet{devore00}, and subsequently adapted to finite volumes by \cite{val12}. We start from the defining relation of the vector potential [$\mathbfit{B}=\nabla\times\mathbfit{A}$] written in spherical coordinates
\begin{eqnarray}
(B_r,B_\theta,B_\phi)=\left( \frac{1}{r\sin \theta} \frac{\partial}{\partial\theta}  \left( \sin\theta A_\phi \right)-\frac{1}{r\sin \theta} \frac{\partial A_\theta}{\partial \phi}, \right. \nonumber \\
\frac{1}{r\sin \theta} \frac{\partial A_r}{\partial\phi} -\frac{1}{r} \frac{\partial}{\partial r} (r A_\phi),\nonumber \\
\left. \frac{1}{r} \frac{\partial}{\partial r} (r A_\theta)-\frac{1}{r} \frac{\partial A_r}{\partial \theta} \right).
\label{curlA} 
\end{eqnarray}
We then employ the DeVore gauge \citep{devore00} in which the radial component of the vector potential is identically zero: $A_r=0$. This same gauge has also been used by \citet{val12}, \citet{amari13}, \citet{moraitis14}, and \citet{yeates16}.

The $\theta$- and $\phi$-components of Equation~\ref{curlA} can then be immediately integrated with the result
\begin{equation}
A_\phi(r,\theta,\phi)=\frac{r_0}{r}A_\phi(r_0,\theta,\phi)-\frac{1}{r}\int_{r_0}^r\mathrm{d}r'\,r'B_\theta(r',\theta,\phi)
\label{aphi} 
\end{equation}
and
\begin{equation}
A_\theta(r,\theta,\phi)=\frac{r_0}{r}A_\theta(r_0,\theta,\phi)+\frac{1}{r}\int_{r_0}^r\mathrm{d}r'\,r'B_\phi(r',\theta,\phi)
\label{atheta} 
\end{equation}
respectively, and $r_0$ is an arbitrary radius inside the volume. These can be written more compactly as
\begin{equation}
\mathbfit{A}(r,\theta,\phi)=\frac{1}{r} \left( r_0\mathbfit{a}(\theta,\phi)+\hat{\mathbfit{r}}\times \int_{r_0}^r\mathrm{d}r'\,r'\mathbfit{B}(r',\theta,\phi)\right) ,
\label{anot0}
\end{equation}
where $\mathbfit{a}(\theta,\phi)=\left( A_\theta(r_0,\theta,\phi),A_\phi(r_0,\theta,\phi) \right)$ is a two-dimensional (2D) integration vector that represents the vector potential on the surface $r=r_0$.

Substituting Equation~\ref{anot0} in the radial component of Equation~\ref{curlA}, and using the divergence-freeness of the field, we get
\begin{equation}
\nabla_\perp\times\mathbfit{a}= \frac{1}{r_0\sin \theta} \left( \frac{\partial}{\partial\theta}  \left( \sin\theta \alpha_\phi \right)-\frac{\partial \alpha_\theta}{\partial \phi}\right) =B_r(r_0,\theta,\phi),
\label{crlanot0}
\end{equation}
where the symbol ``$\perp$'' stands for the normal direction to the radial one. This relation can be thought of as a restriction on the possible choices for the remaining freedom in the gauge of $\mathbfit{A}$. 

\subsubsection{DeVore Simple Gauge}

A first simple solution to Equation~\ref{crlanot0}, which corresponds to a first family of gauge, is given by the relations
\begin{equation}
\alpha_\phi(\theta,\phi)= \frac{cr_0}{\sin \theta} \int_{\theta_0}^\theta\mathrm{d}\theta'\,\sin \theta' B_r(r_0,\theta',\phi)
\label{aphi0} 
\end{equation}
and
\begin{equation}
\alpha_\theta(\theta,\phi)= -(1-c)r_0\sin \theta \int_{\phi_0}^\phi\mathrm{d}\phi'\,B_r(r_0,\theta,\phi'),
\label{atheta0} 
\end{equation}
where $\theta_0$, $\phi_0$ are arbitrary angles, and $c\in [0,1]$ a constant, typically $c=1/2$. The vector potential in this gauge is given by Equations~\ref{anot0}, \ref{aphi0}, and \ref{atheta0}, and its calculation requires the computation of four one-dimensional integrals. In our present implementation, the integrations are treated with the simple trapezoidal method, although more sophisticated methods can be envisioned. In the following we refer to this gauge family as the DeVore simple gauge (DVS).

\subsubsection{DeVore Coulomb Gauge}

An alternative family of gauges can be obtained when $\mathbfit{a}$, the vector potential on the surface $r=r_0$, satisfies the Coulomb gauge, $\nabla_\perp\cdot\mathbfit{a}=0$. This is automatically fulfilled if $\mathbfit{a}$ is expressed through a 2D scalar function [$u$], as given by
\begin{equation}
\mathbfit{a}=\hat{\mathbfit{r}}\times\nabla_\perp u=\frac{1}{r_0}\left( -\frac{1}{\sin\theta}\frac{\partial u}{\partial\phi},\frac{\partial u}{\partial\theta}\right) .
\label{ata0u} 
\end{equation}
By substituting Equation~\ref{ata0u} in Equation~\ref{crlanot0} we find that the function $u$ satisfies the 2D Poisson equation $\nabla_\perp^2u=B_r(r_0,\theta,\phi)$. This equation can be solved easily with the additional assumption of the Dirichlet boundary condition, $u=0$, along the boundary of the $r=r_0$ plane. The use of this boundary condition allows for a more generic $B_r$-distribution, such as a non flux-balanced one, compared to the corresponding Neumann boundary condition, as follows from the uniqueness condition of the Poisson equation. In our particular implementation we solve the Poisson equation with the aid of a \textsf{FORTRAN} routine from the \textsf{FISHPACK} library \citep{ss79}. 

The vector potential in this gauge family is given by Equations~\ref{anot0}, and \ref{ata0u}, and its calculation requires the computation of two one-dimensional integrals, and of a 2D Poisson equation. While \cite{amari13} named this gauge the restricted DeVore gauge, we will denote it as the DeVore Coulomb gauge (DVC), because, when applied to the potential field, this satisfies the Coulomb gauge in the whole volume \citep[see Section~2.2 in][]{val12}.

\subsubsection{Reference Boundary Choice}

Apart from the two gauges for the integration vector (DVS and DVC) there is also the choice of the location of the reference surface in the vector potential calculation. We consider here only the two interesting cases, the bottom and the top surfaces of the volume, similarly to Equations~10 and 11 of \citet{val12}, although other choices exist also \citep[see Section~2.2.3 of][]{valori16}. These cases are denoted by the letters ``b'' and ``t'', respectively, and they follow the symbol of the gauge. A given vector potential can thus be in any of the four different gauges: DVSb, DVSt, DVCb, or DVCt.

The vector potential for the potential field can be obtained with the same method by just replacing $\mathbfit{B}$ with $\mathbfit{B}_\mathrm{p}$ in the above relations. It can thus also be in four different gauges and since in general the gauges of the two vector potentials are independent, there are 16 possible combinations in the calculation of helicity.

\section{Validation}
     \label{S-validation}

\subsection{Test Datasets}

The performance of the helicity calculation method is checked against the semi-analytical, force-free field solutions of \citet[][hereafter LL fields]{lowlou90}. We simulate an active region (AR) with linear dimensions $\approx 200-250$~Mm at a solar latitude of $\pm 30^\circ$, which translates to the angular dimensions $\approx 15-20^\circ$ on the Sun. The particular location of the synthetic AR is of course irrelevant in the computation of helicity. The precision of our method is only very marginally affected by the particular values of the domain. Nonetheless, we wish to work with a synthetic AR domain as close as possible to a realistic case. For this we assume the spherical wedge volume $V=\{(r,\theta,\phi): r\in[700\,{\rm Mm},900\,{\rm Mm}],\theta\in[50^\circ,70^\circ],\phi\in[10^\circ,30^\circ]\}$ with height $\approx 200$~Mm and the AR at the small side of the wedge. The volume is discretized by a uniform grid of size $129\times129\times129$ grid points. 
 
\begin{figure}
\centering
\includegraphics[width=0.85\textwidth]{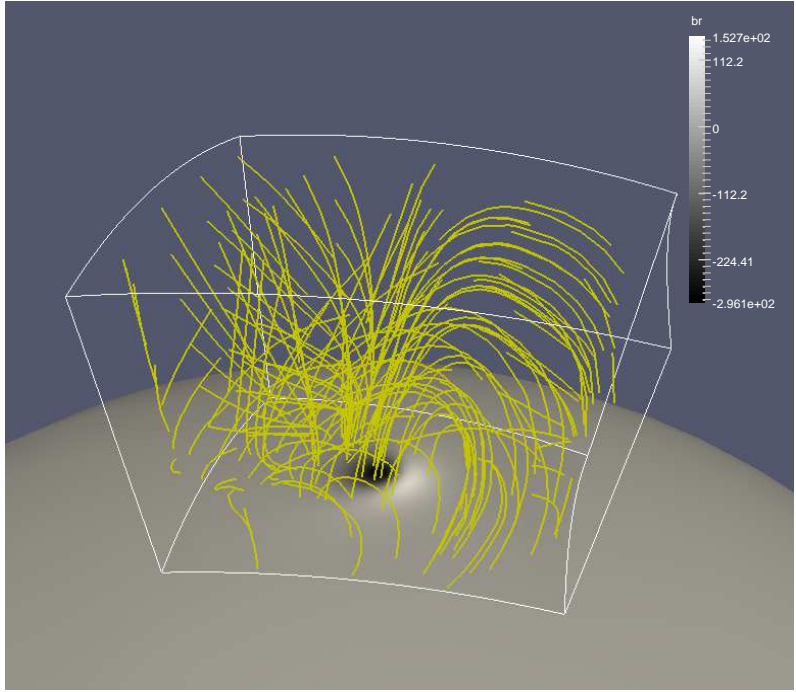}
\caption{Plot of the 129$^3$-grid LL magnetic field that was used in the tests. Shown are the image of the radial component of the LL field on the photosphere with the designated color scale and angular dimensions 20$^{\rm o}\times$20$^{\rm o}$, and a set of representative field lines up to the height of 200~Mm from the solar surface.}
\label{llplot}
\end{figure}

The usual parameters of the LL fields are also assumed, $n=m=1$, while the source is placed at a depth of $\approx 30$~Mm below the AR (or $l=0.3$ in LL notation), and is rotated by the angle $\phi=\pi/4$ with respect to the radial direction. A plot of the LL field, denoted hereafter as $\mathbfit{B}_{\rm LL}$, with some representative field lines is shown in Figure~\ref{llplot}. In order to check the effect of resolution we also use the same volume, but discretized by $257\times257\times257$ grid points.

The first step in estimating the helicity of $\mathbfit{B}_{\rm LL}$ is to calculate its scalar potential $\Phi_{\rm LL}$ as described in Section~\ref{S-helicity}. The corresponding potential field is obtained from the relation $\mathbfit{B}_{\rm p,LL}=\nabla\Phi_{\rm LL}$, where the numerical derivatives used are of second order inside $V$, and on the boundary are given by Equation~\ref{batbound}. The solenoidality of these magnetic fields is verified below.

\subsection{Solenoidality Verification}

The divergence-freeness of a magnetic field is quantified in three ways. The first is by the average fractional flux increase [$\left\langle |f_i|\right\rangle$] defined by \cite{wheatland00}.
 
The second verification metric is the flux imbalance ratio [$\epsilon_{\rm flux}$]. This is defined by 
\begin{equation}
\epsilon_{\rm flux}=\frac{|\Phi^+-\Phi^-|}{\Phi^++\Phi^-}
\end{equation}
where $\Phi^+$ and $\Phi^-$ are the positively-defined fluxes entering and leaving the volume through all of its boundaries, respectively. A highly flux-balanced field, indicating also a solenoidal field, corresponds to the limit $\epsilon_{\rm flux}=0$, while the opposite happens for $\epsilon_{\rm flux}=1$. 

The third metric used relates to the magnetic energies. For each $\mathbfit{B}_{\rm LL}$ and $\mathbfit{B}_{\rm p,LL}$ at each resolution, we calculate its magnetic energy (in arbitrary units) as $E=\int B^2\,\mathrm{d}V$ (\textit{cf.} Table~\ref{tab1}). By substracting the magnetic energy of the potential field from the magnetic energy of the LL field, we get the free energy: $E_{\rm c}$. Table~\ref{tab1} indicates that the free energy corresponds to $\approx 26\,\%$ of the total energy of the LL field, a value typical for LL fields, independently of the resolution. The solenoidality can finally be estimated thanks to the energy ratio $E_{\rm div}/E$, with $E_{\rm div}$ a pseudo-energy given by
\begin{equation}
E_{\rm div}=\left|2 \int_V \mathbfit{B}_{\rm p,LL}\cdot(\mathbfit{B}_{\rm LL}-\mathbfit{B}_{\rm p,LL})\,\mathrm{d}V\right|.
\end{equation}
This energy vanishes for perfectly solenoidal fields and quantifies the violation of Thompson's theorem by non-purely solenoidal fields \citep{val13}. This quantity is more widely used to estimate non-solenoidality \citep{pariat15,valori16,pariat17,polito17}. Table~\ref{tab1} shows that the small values of $E_{\rm div}/E$ is again an indication of the good level of solenoidality and another proof of the validity of the LL field as a test field, since nonsolenoidality was shown to be the strongest source of errors in helicity estimations \citep[see][]{valori16}.

\begin{table}
\caption{Parameters for the input LL fields, and their corresponding potential fields.}
\begin{tabular}{llccccc}
\toprule
grid & field & $10^{4}\,   \left\langle |f_i| \right\rangle $ & $10^{3}\,  \epsilon_{\rm flux}$ & $E$ & $E_{\rm c}/E$ & $10^{3}\,  E_{\rm div}/E$ \\
\midrule
\multirow{2}{*}{$129^3$} & $\mathbfit{B}_{\rm LL}$ &  $2.21$ & $1.70$ & 45.3 & \multirow{2}{*}{0.262} & \multirow{2}{*}{$1.10$} \\
 & $\mathbfit{B}_{\rm p,LL}$ & $1.15$ & $1.83$ & 33.4 &  & \\
\midrule
\multirow{2}{*}{$257^3$} &$\mathbfit{B}_{\rm LL}$ &  $2.16$ & $2.15$ & 45.2 & \multirow{2}{*}{0.261} & \multirow{2}{*}{$2.51$} \\
 & $\mathbfit{B}_{\rm p,LL}$ & $2.14$ & $2.23$ & 33.4 &  & \\
\bottomrule
\end{tabular}
\label{tab1}
\end{table}

\subsection{Vector-Potentials Verification}

The second main step in the volume-helicity calculation is the computation of the vector potentials. It is important to check the quality of the reconstruction of the vector potential, \textit{i.e.} that the curl of the computed vector potential corresponds indeed to its source magnetic field. For this, we use two methods. 

First, we directly compare the $r$, $\theta$, and $\phi$ components of the original fields, $\mathbfit{B}_{\rm LL}$ and $\mathbfit{B}_{\rm p,LL}$, with their respective vector-potential-reconstructed magnetic fields, $\nabla\times \mathbfit{A}_{\rm LL}$ and $\nabla\times \mathbfit{A}_{\rm p,LL}$. To compare the components we compute their linear (Pearson) correlation coefficients, which are presented in Table~\ref{tab2}.
 
In addition, we compute the vector fields'` comparison metrics given by \cite{schrijver06}, namely the vector correlation [$C_{\rm vec}$], the Cauchy--Schwarz metric [$C_{\rm CS}$], the complement of the normalized vector error [$E_{\rm n}'$], the complement of the mean vector error [$E_{\rm m}'$], and the total energy normalized to that of the input field [$\epsilon$]. For two arbitrary vector fields, original [$\mathbfit{X}$], and reconstructed [$\mathbfit{Y}$], consisting of $N$ points, these read
\begin{eqnarray}
C_{\rm vec}=&\frac{{\displaystyle \sum_i \mathbfit{X_i}\cdot \mathbfit{Y_i}}}{\left( {\displaystyle \sum_i |\mathbfit{X_i}|^2}{\displaystyle \sum_i |\mathbfit{Y_i}|^2} \right)^{1/2}}\\
C_{\rm CS}=&\frac{1}{N}{\displaystyle \sum_i \frac{\mathbfit{X_i}\cdot \mathbfit{Y_i}}{|\mathbfit{X_i}| |\mathbfit{Y_i}|}}\\
E_{\rm n}'=&1-\frac{{\displaystyle \sum_i |\mathbfit{X_i}-\mathbfit{Y_i}|}}{{\displaystyle \sum_i |\mathbfit{X_i}|}}\\
E_{\rm m}'=&1-\frac{1}{N}{\displaystyle \sum_i \frac{|\mathbfit{X_i}-\mathbfit{Y_i}|}{|\mathbfit{X_i}|}}\\
\epsilon=&\frac{{\displaystyle \sum_i |\mathbfit{Y_i}|^2}}{{\displaystyle \sum_i |\mathbfit{X_i}|^2}}
\end{eqnarray}
All metrics indicate the good agreement of the two fields by values close to unity.

In Table~\ref{tab2} we present the values of these parameters for a few example reconstructions of the LL field, and separately, of the potential field. We performed different tests using: the two different gauges discussed in Section~\ref{S-helicity-vp}, DVS and DVC; the two different spatial resolutions, 129$^3$ and 257$^3$; and/or the two different reference surfaces: bottom and top. As an example, consider the first row in Table~\ref{tab2}, where the values correspond to the reconstruction of the LL field of $129^3$ grid points with a vector potential in the simple DeVore gauge, and with the top surface as the reference surface for the integration. 

Table~\ref{tab2} demonstrates that the correlation coefficients are all very close to 1. The value for the $r$-component presents the weakest correlation since this component involves the most numerical operations. The values for Schrijver's metrics also indicate that the reconstruction is excellent in all the cases. We further note that the reconstruction in spherical coordinates has comparable accuracy to the corresponding one for the Cartesian case \citep[Table 8 in][]{valori16}.
 
\begin{table}
\caption{Metrics for the reconstruction of the magnetic field from the respective vector potential.}
\resizebox{0.98\textwidth}{!}{
\begin{tabular}{ccccccccccc}
\toprule
& &  & \multicolumn{3}{c}{correlation coefficients of} &  \\
& &  & \multicolumn{3}{c}{$\mathbfit{B}$ \textit{vs.} $\nabla \times \mathbfit{A}$} & \multicolumn{5}{c}{Schrijver metrics} \\
\cline{4-11}
field & gauge & grid & $B_r$ & $B_\theta$ & $B_\phi$ & $C_{\rm vec}$ & $C_{\rm CS} $ & $E_{\rm n}'$ & $E_{\rm m}'$ & $\epsilon$ \\
\midrule
\multirow{4}{*}{$\mathbfit{B}_{\rm LL}$}   & DVSt & $129^3$ & 0.9999 & 1.0000 & 1.0000 & 0.9999 & 1.0000 & 0.9948 & 0.9959 & 0.9980 \\
   & DVSt & $257^3$ & 0.9999 & 1.0000 & 1.0000 & 0.9999 & 1.0000 & 0.9942 & 0.9949 & 0.9986 \\
   & DVSb & $129^3$ & 0.9990 & 1.0000 & 1.0000 & 0.9995 & 0.9986 & 0.9814 & 0.9613 & 1.0025 \\
   & DVCt & $129^3$ & 0.9999 & 1.0000 & 1.0000 & 0.9999 & 0.9999 & 0.9947 & 0.9953 & 0.9980 \\
\midrule
\multirow{4}{*}{$\mathbfit{B}_{\rm p,LL}$} & DVSt & $129^3$ & 1.0000 & 1.0000 & 1.0000 & 1.0000 & 0.9998 & 0.9888 & 0.9829 & 0.9977 \\
   & DVSt & $257^3$ & 0.9995 & 1.0000 & 1.0000 & 0.9997 & 0.9962 & 0.9570 & 0.9288 & 0.9990 \\
   & DVSb & $129^3$ & 0.9999 & 1.0000 & 1.0000 & 0.9999 & 0.9978 & 0.9843 & 0.9627 & 1.0008 \\
   & DVCt & $129^3$ & 1.0000 & 1.0000 & 1.0000 & 1.0000 & 0.9997 & 0.9888 & 0.9824 & 0.9977 \\
\bottomrule
\end{tabular} }
\label{tab2}
\end{table}

From the values of Table~\ref{tab2} there can be drawn a few conclusions. First, the differences for the two different  spatial resolutions are very small for both $\mathbfit{B}_{\rm LL}$, and $\mathbfit{B}_{\rm p,LL}$, with the latter exhibiting slightly better metrics in the lower-resolution field. Second, the differences between the reconstructions performed with the DVS or the DVC gauges are practically nonexistent, for both $\mathbfit{B}_{\rm LL}$ and $\mathbfit{B}_{\rm p,LL}$. We note here that the vector potentials in the DVC gauge satisfy the Coulomb gauge on the surface $r=r_0$ to a large degree. This can be seen by taking the $129^3$-grid potential field as an example, where the absolute fractional flux increase of its vector potential in the DVCt gauge is $\left\langle |f_i|\right\rangle =1.4\times 10^{-7}$, a value comparable to the best-performing methods of \citet[][Figure~7f]{valori16}.

The most important differences in the reconstruction metrics are found, however, between the top and bottom reference surfaces, mostly in the parameters $E_{\rm n}'$ and $E_{\rm m}'$, which are the most sensitive ones. One can note that using the top boundary yields more precise results. This is a general characteristic of the helicity-computation methods based on the DeVore gauge and the properties of the data tested, as already noted by \citet{moraitis14,pariat15,valori16}. In solar-like datasets the magnetic field at the top surface is weaker and smoother than in the bottom one, and, as a result, the integrations involved in the vector-potential computation that start from the top surface are also smoother.

Additionally, the values of the metrics for the potential fields [$\mathbfit{B}_{\rm p,LL}$] are, in general, slightly inferior to the respective ones for $\mathbfit{B}_{\rm LL}$. This is to be expected since the potential field is derived from the LL one and thus carries any numerical errors in it.

As a final check we compute relative helicity as given by Equation~\ref{helc} using all of the possible different gauge combinations for the vector potentials of $\mathbfit{B}_{\rm LL}$ and its potential fields. The produced helicity values for the $129^3$-grid fields are all very similar, ranging in $-145.2\pm 0.4$ (in arbitrary units), indicating that helicity is indeed independent of the chosen gauge up to a factor $2\times 10^{-3}$. 

As a conclusion of this Section we can say that the developed helicity computation method in spherical coordinates is performing extremely well. This is deduced from the high solenoidality of the constructed potential magnetic fields, and also, from the level of reproduction of the magnetic fields from the respective vector potentials. In addition, the method is computationally very efficient, as it requires only a few minutes to compute helicity for the $257^3$-grid field on a commercial laptop.

\section{Application to Solar Active Region 3D Extrapolated Magnetic Fields}
     \label{S-application} 

Having established that our method performs as expected on a semi-analytical magnetic field, we now apply it to typical magnetic-field data from a reconstructed solar active region. This will show the performance of our method to a more solar-like case, and also, it will give the opportunity to compare our results with more approximate helicity calculations methods.

\subsection{NLFF Field Model}
     \label{Ss-data} 
     
We use a set of data-constrained non-linear force-free field models of the NOAA AR 11060 (\textsf{SOL2010-04-08T02:35:00L110C176}). These have been constructed in order to topologically study the specific active region \citep{savcheva16}. All computations were performed in spherical coordinates.

The fields were produced by the flux-rope-insertion method \citep{vanB04}, which involves various intermediate steps that we briefly summarise here. First, a global potential magnetic field is computed from a low-resolution synoptic magnetogram. Then, a modified potential-field extrapolation is performed starting from a high-resolution magnetogram centred on the active region, and with lateral boundary conditions given by the global field. The relevant calculations are done directly with the vector potential [$\mathbfit{A}_\mathrm{p,MF}$], to ensure solenoidality of the respective magnetic field. This potential field [$\mathbfit{B}_\mathrm{p,MF}=\nabla\times\mathbfit{A}_\mathrm{p,MF}$], is then modified with two sources in the photosphere where the inserted flux rope is to be anchored. 

The insertion of the flux rope is done by modifying the initial vector potential with a combination of axial and poloidal flux. The model is then relaxed towards a force-free state using the magnetofrictional (MF) method. In the MF code, it is the vector potential again that is relaxed, $\mathbfit{A}_{\rm MF}$, and not the magnetic field $\mathbfit{B}_{\rm MF}=\nabla\times\mathbfit{A}_{\rm MF}$. During the MF relaxation, $\hat{\mathbfit{n}}\cdot\mathbfit{B}_{\rm MF}$ is held constant at the photospheric boundary; the other five boundaries are very far away from the AR core so that the field there is very well approximated by the potential one. The potential magnetic field and its respective vector potential remain constant and equal to their value at the initial instant during the MF process, as this is not a temporal evolution. The full details of the method used to produce the NLFF fields are described by \cite{savcheva16} and references therein.

A magnetic-field model is considered as stable if, after the relaxation, it has reached a force-free equilibrium where no residual Lorentz force is present, otherwise it is considered to be marginally stable or unstable. An unstable model is obtained from a stable model by adding slightly more axial flux to the stable inserted-flux-rope parameters. The parameters used in the present study are the same as those used by \citet{savcheva16}. 

\begin{figure}
\centering
\begin{minipage}[b]{0.5\textwidth}
\centering
\includegraphics[width=\textwidth,trim=25 0 20 0,clip]{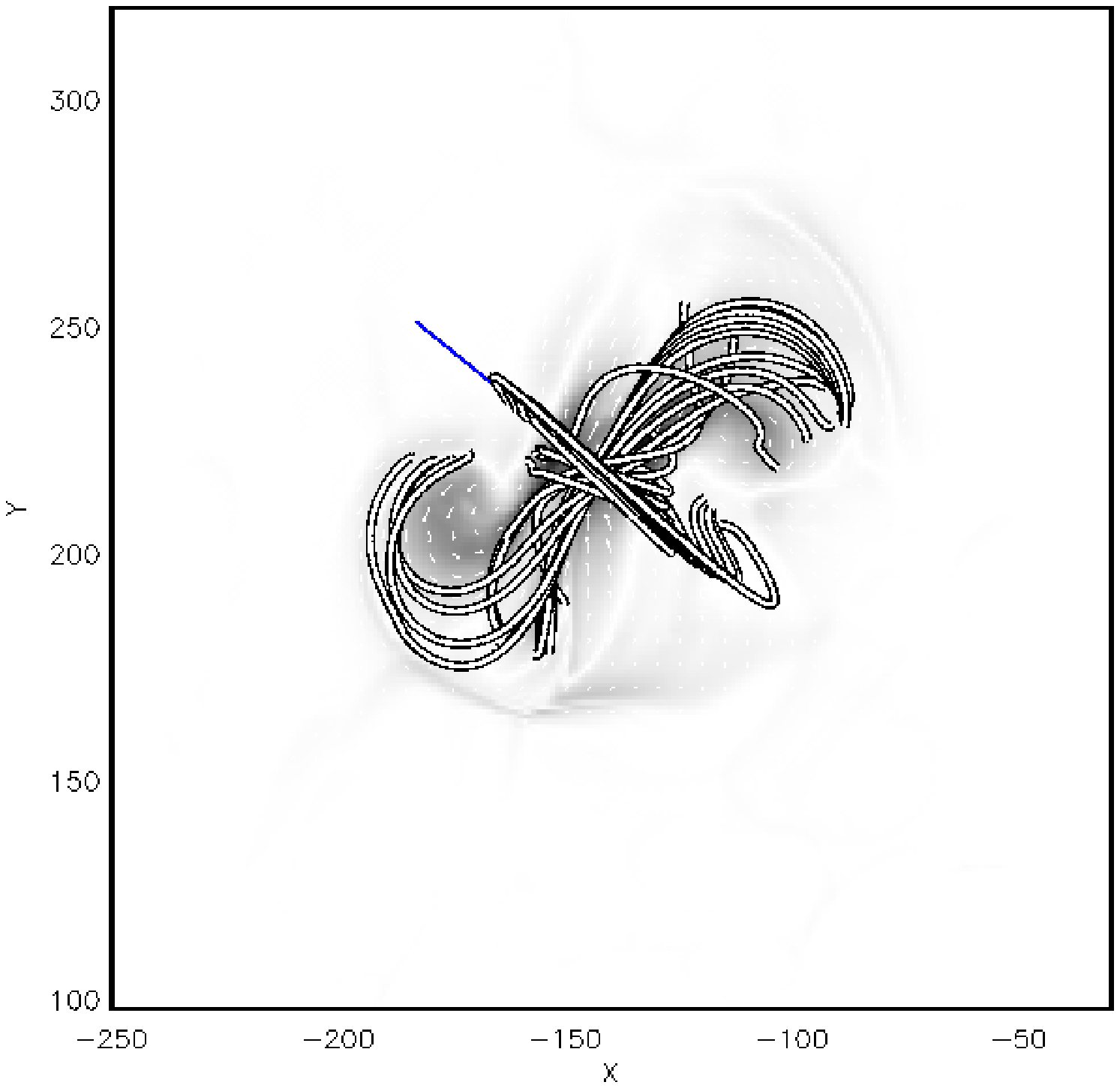}
\end{minipage}%
\begin{minipage}[b]{0.5\textwidth}
\centering
\includegraphics[width=\textwidth,trim=0 30 0 70,clip]{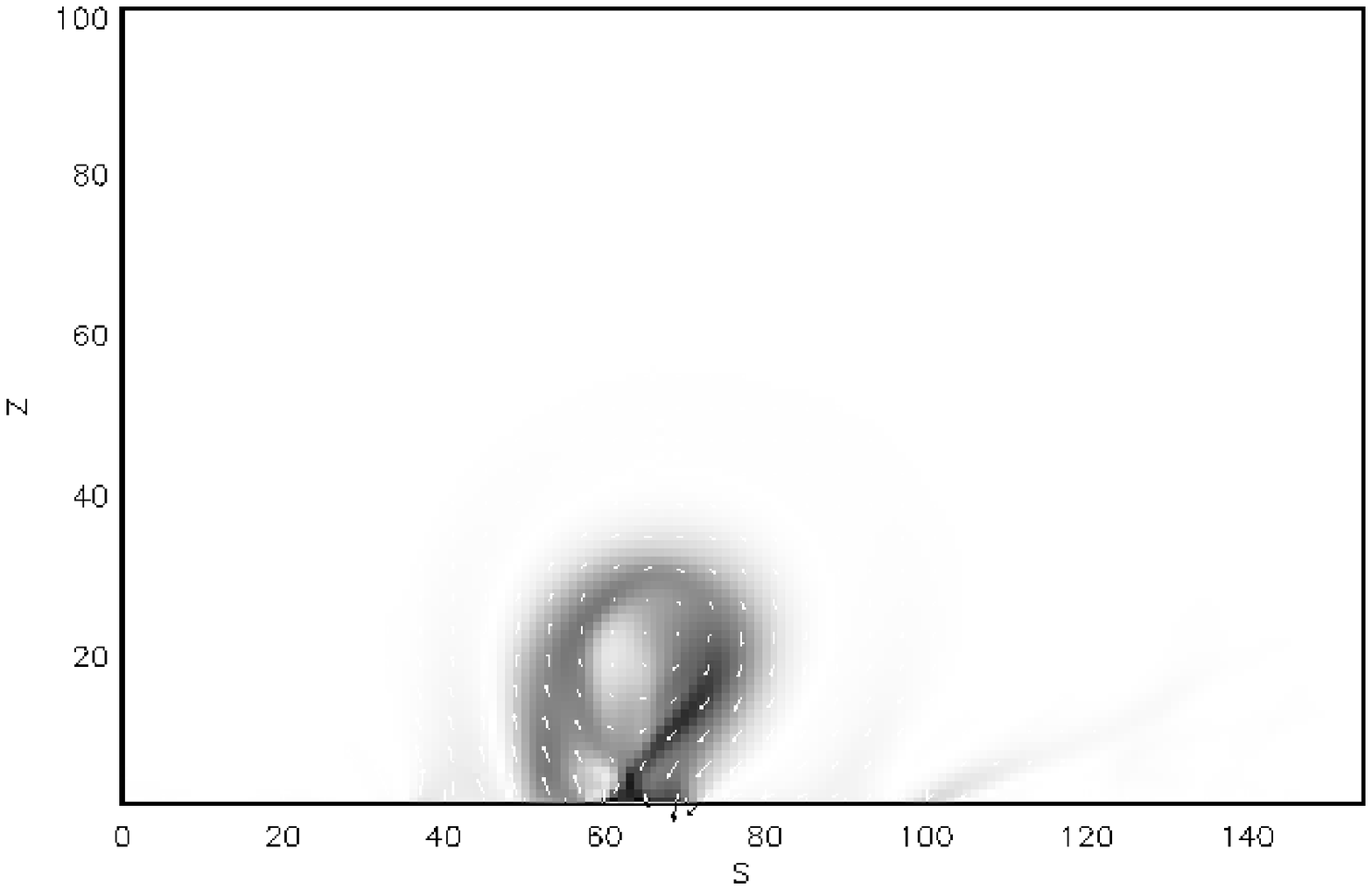}
\end{minipage}\\
\begin{minipage}[b]{0.5\textwidth}
\centering
\includegraphics[width=\textwidth,trim=25 0 20 0,clip]{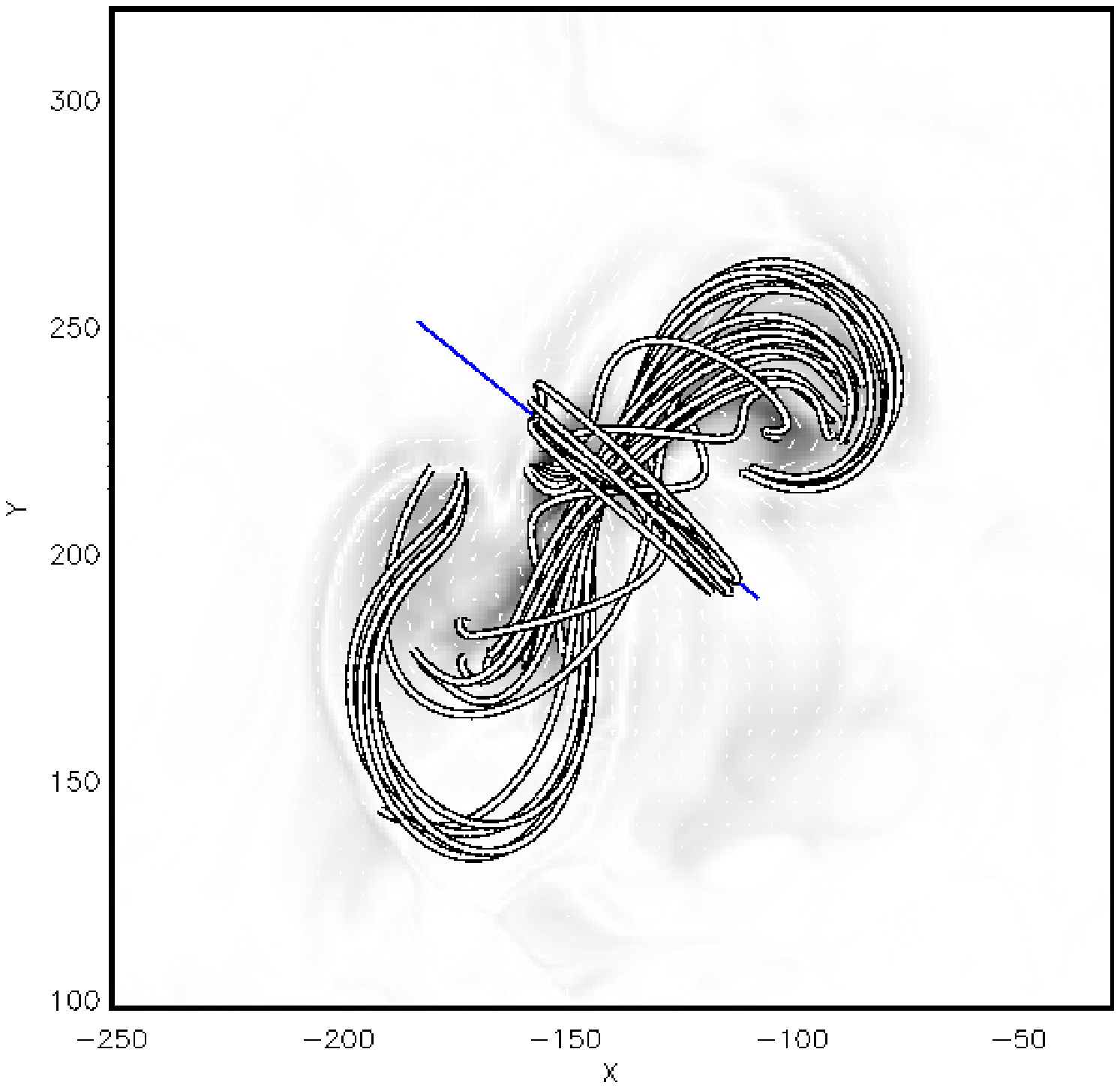}
\end{minipage}%
\begin{minipage}[b]{0.5\textwidth}
\centering
\includegraphics[width=\textwidth,trim=0 30 0 70,clip]{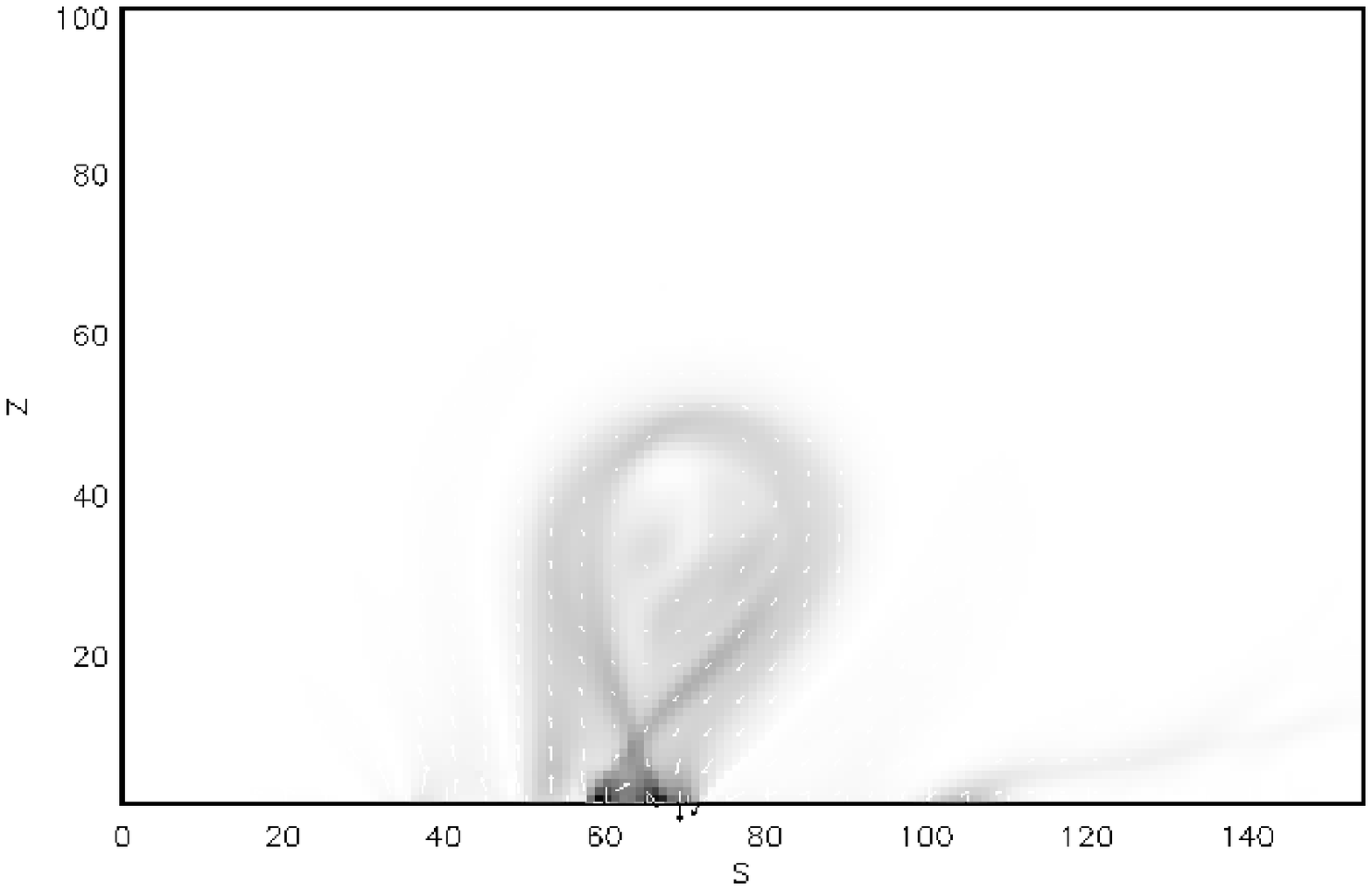}
\end{minipage}\\
\begin{minipage}[b]{0.5\textwidth}
\centering
\includegraphics[width=\textwidth,trim=25 0 20 0,clip]{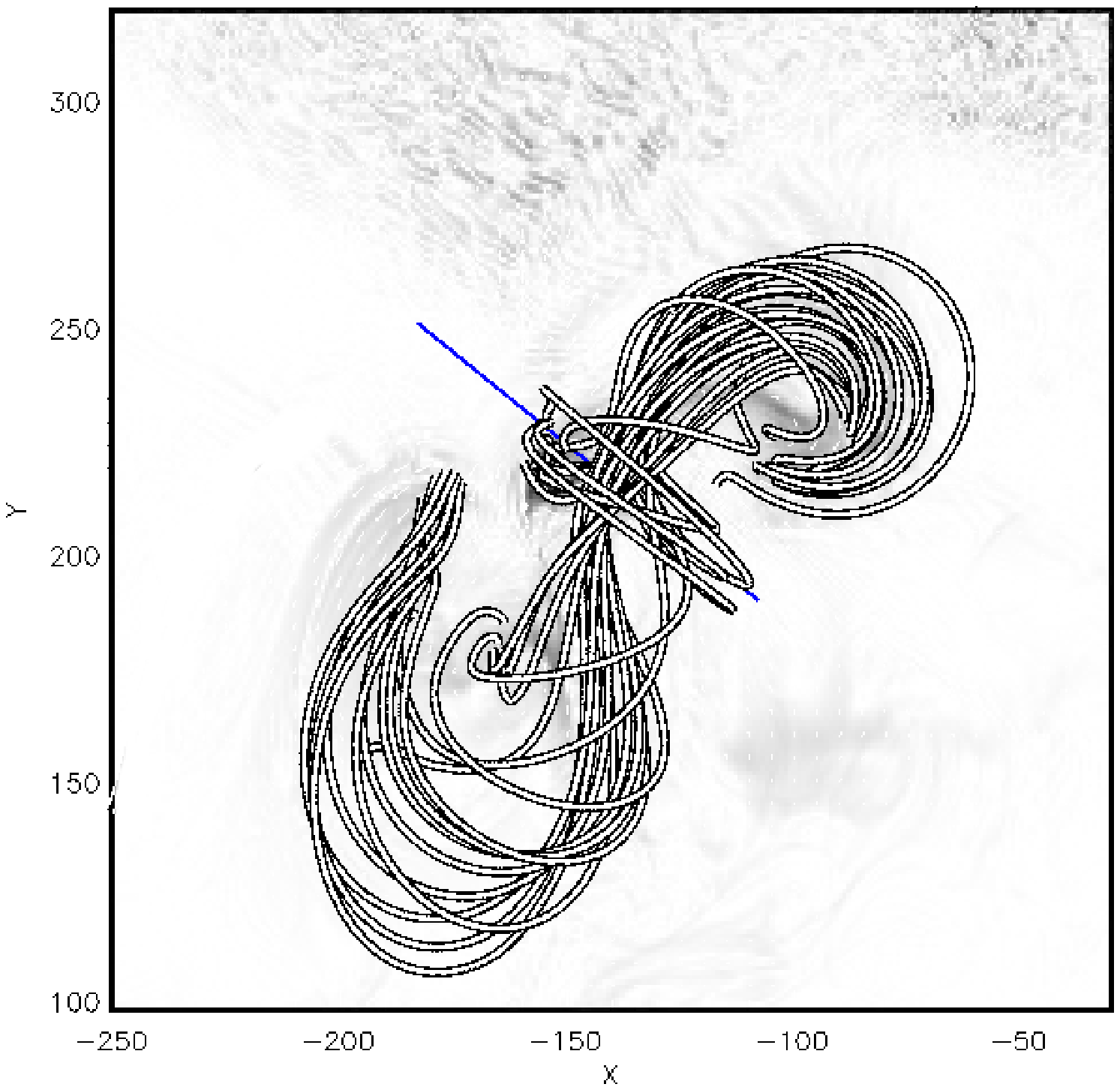}
\end{minipage}%
\begin{minipage}[b]{0.5\textwidth}
\centering
\includegraphics[width=\textwidth,trim=0 30 0 70,clip]{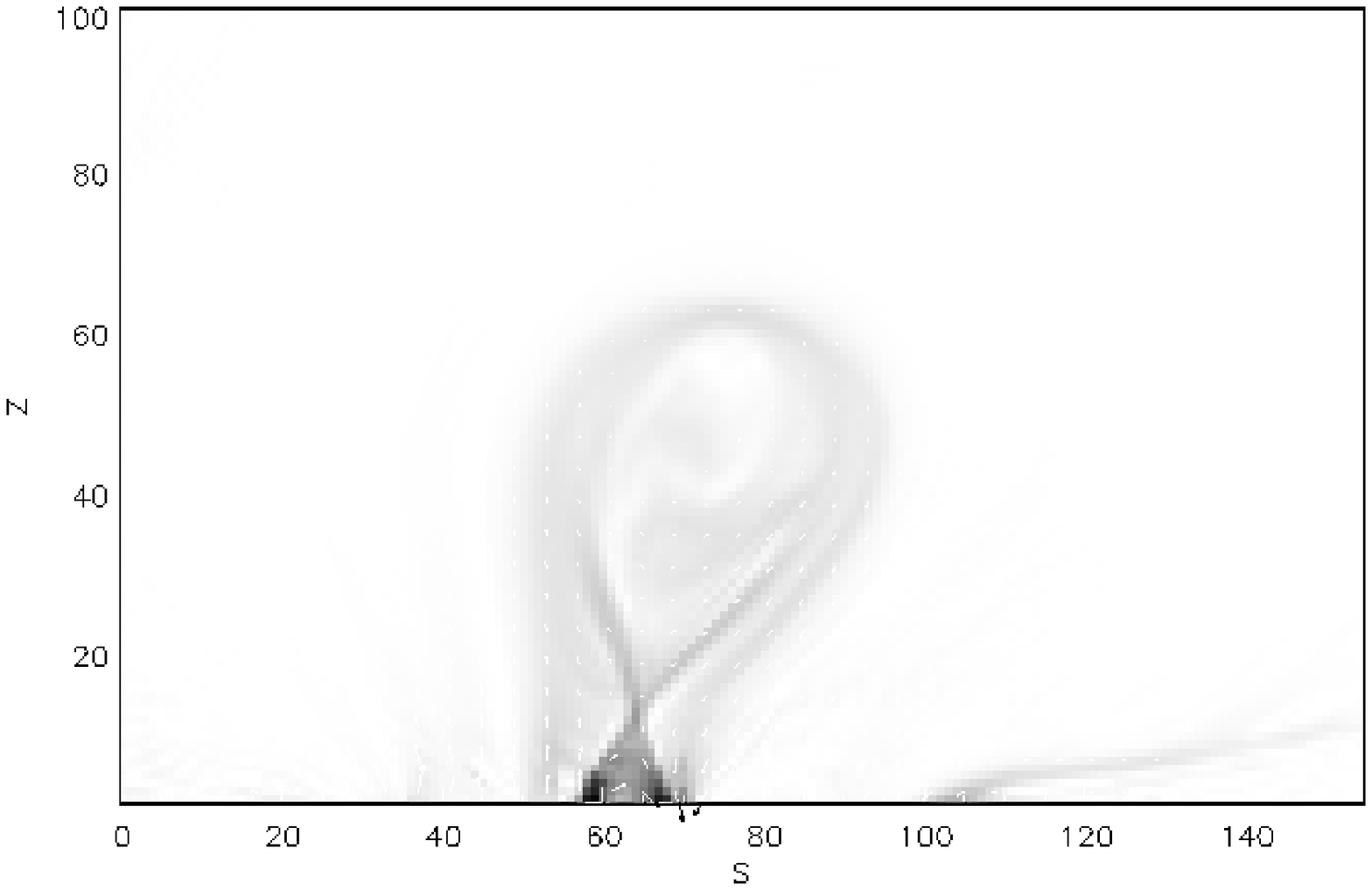}
\end{minipage}
\caption{Evolution of the flux-rope morphology with iteration number. For the snapshots at $25,000$ (top), $90,000$ (middle), and $155,000$ (bottom) iterations, the plots show: a number of characteristic field lines overplotted on the horizontal cut of the current density at $z=10$~Mm (left), and the vertical cut of the current density along the blue line of the left plot (right). Spatial scales are in units of Mm.}
\label{magfld}
\end{figure}

This marginally unstable NLFF model leads also to an ``eruption'' resembling a coronal mass ejection during the MF evolution of the field, as was demonstrated by \citet{savcheva16}. The gradual expansion and elevation of the flux rope can also be inferred from the evolution of the field-line connectivity and of the flux-rope height that are shown in Figure~\ref{magfld}. A data-constrained MHD simulation starting from a similar initial condition and proving that an unstable model obtained with the flux-rope insertion method is unstable as well in the MHD sense was performed by \citet{kliem13} and demonstrated a good correspondence with the observations.

We use 38 snapshots of the magnetic-field evolution, three at very early stages at 0, 100, and $1000$ iterations, and the remaining from $5000$ to $175,000$ iterations with a step of $5000$ iterations. The magnetic field covers an area of $\approx$40$^\circ\times$40$^\circ$ in the $\theta$--$\phi$ plane, and it extends 805~Mm in the radial direction. The magnetic-field datacubes have the dimensions of 385$\times$385$\times$385 grid points. The grid in the $\phi$- direction is uniform with a stepsize of $0.11^\circ$, while along the $r$-, and $\theta$-directions it is nonuniform. In the $\theta$-direction the stepsize increases from $0.08^\circ$ to $0.11^\circ$ as we go to the Equator, while along $r$ the stepsize increases from 1.4~Mm in the photosphere, to 3~Mm in the uppermost level. The grid that we use here is thus structured.

\subsection{Results}
     \label{Ss-results} 

In this Section we compute the instantaneous value of helicity during the relaxation of the magnetic field with the magnetofrictional method, using the method of this article, and also, the approximate method based on Equation~\ref{helb}. The method of this article will be denoted as the exact method since it makes no assumptions, it treats appropriately all boundaries, and it performs very well with the LL field, as was shown in the previous section. 

\subsubsection{Exact Method}

To compute magnetic helicity with the method of this article we start from the magnetic field [$\mathbfit{B}_{\rm MF}$] that is defined on a nonuniform grid, and we interpolate it to a uniform grid on the same volume. This is not a necessary step in general, but the current implementation of the method is more accurate with a uniform grid. A simple trilinear interpolation is sufficient to obtain similar levels of accuracy as in the LL case, as it will shortly be seen. 

\begin{figure}
\centering
\includegraphics[width=0.8\textwidth]{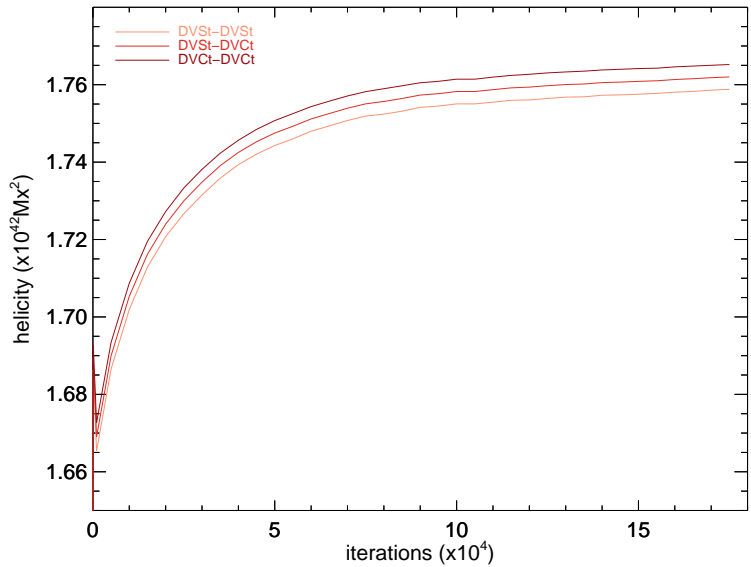}
\caption{Evolution of relative magnetic helicity [$H$] during the magnetofrictional relaxation as computed with the method of this article, Equation~\ref{helca0}. Computations are done with three different gauge combinations for the vector potentials of the original and potential magnetic fields, shown in that order in the label.}
\label{helsava}
\end{figure}

From the interpolated magnetic field [$\mathbfit{B}_{\rm DV}$] we compute the vector fields, $\mathbfit{B}_{\rm p,DV}$, $\mathbfit{A}_{\rm DV}$, and $\mathbfit{A}_{\rm p,DV}$, using the method described in Section~\ref{S-helicity}. Note that the potential field is computed at each snapshot in order to take into account possible changes of $\hat{\mathbfit{n}}\cdot\mathbfit{B}_{\rm DV}$ caused by the interpolation from $\mathbfit{B}_{\rm MF}$. The helicity computation that uses the method of this article, is thus based on the relation
\begin{equation}
H=\int_V\,\mathrm{d}V (\mathbfit{A}_{\rm DV}+\mathbfit{A}_\mathrm{p,DV})\cdot (\mathbfit{B}_{\rm DV}-\mathbfit{B}_\mathrm{p,DV}).
\label{helca0}
\end{equation}
The evolution of magnetic helicity as computed with the exact method is shown in Figure~\ref{helsava}. The helicity obtained is an increasing function of iteration number, after the first couple of snapshots of the flux-rope insertion, and it seems to saturate at a value a little above $\approx 1.76\times10^{42}\,{\rm Mx}^2$. The increasing helicity evolution pattern is in contrast to the respective free energy one, which is decreasing as shown in Figure~8 of \citet{savcheva16}, and also, to the general trend between free energy and helicity \citep{tziotziou14}. Nevertheless, it is in line with the removal of negative helicity as the flux rope expands and subsequently erupts during the MF evolution.

The level of gauge invariance is quite high, as can be deduced from the values of helicity computed with three different combinations of the DVSt and DVCt gauges for the vector potentials of the original and of the potential magnetic fields. The three different helicity curves of Figure~\ref{helsava} practically coincide, exhibiting differences $\lesssim 0.2\,\%$. 

The computed potential magnetic field is sufficiently divergence-free, as the average (over all snapshots) values for the metrics $\left\langle |f_i|\right\rangle =3.0\times10^{-4}$ and $\epsilon_{\rm flux} =3.5\times10^{-4}$ indicate. The average values of the reconstruction metrics for the vector potential $\mathbfit{A}_{\rm DV}$ (as in Table~\ref{tab2}) for this case are all $>0.99$, except $E_{\rm n}'=0.97$, $E_{\rm m}'=0.97$, and $\epsilon=0.97$. Similarly, for $\mathbfit{A}_{\rm p,DV}$ they are $>0.98$, except $E_{\rm n}'=0.93$, $E_{\rm m}'=0.90$, and $\epsilon=1.03$. The vector-potential reconstructions are thus slightly inferior compared to those of the LL case, but still very good. Combined with the fact that the computed helicity is also gauge-independent to a large degree, as Figure~\ref{helsava} shows, this leads to the conclusion that this can be considered as the actual AR helicity, and it further justifies the characterization of the method as exact.

\subsubsection{Approximate Method}

The second method that we use to compute helicity follows the reasoning of \citet{bobra08}, that is, it employs the definition of Equation~\ref{helb}. We furthermore use two different sets of vector fields in the computations. The first is the direct outcome of the magnetofrictional code. The relation that is used in this case is
\begin{equation}
H_R=\int_V\,\mathrm{d}V (\mathbfit{A}_{\rm MF}\cdot \mathbfit{B}_{\rm MF} - \mathbfit{A}_\mathrm{p,MF}\cdot \mathbfit{B}_\mathrm{p,MF})+\int_S\,\mathrm{d}S \chi_\mathrm{MF} B_{{\rm MF},r},
\label{helb1}
\end{equation}
where all MF-related quantites are explicitly stated. The scalar quantity $\chi_{\rm MF}$ at a point $\mathcal{P}$ of the photospheric plane $S$ is given by the line integral $\chi_{\rm MF}(\mathcal{P})=\int_\mathcal{O}^\mathcal{P} \mathrm{d}\mathbfit{l}\cdot(\mathbfit{A}_{\rm MF}-\mathbfit{A}_\mathrm{p,MF})$, where $\mathcal{O}$ is a reference point on the same plane. Notice also that the vector fields involved in Equation~\ref{helb1} are the original, uninterpolated ones, and that the potential field and its respective vector potential do not change during the relaxation. The MF code evolves $\mathbfit{A}_\mathrm{MF}$ but on the (coronal) boundary the condition $\mathbfit{A}_\mathrm{MF}\times\mathbfit{A}_\mathrm{p,MF}=0$ is maintained valid at all times so that Equation~\ref{helb1} is as accurate as possible.

\begin{figure}
\centering
\includegraphics[width=0.8\textwidth]{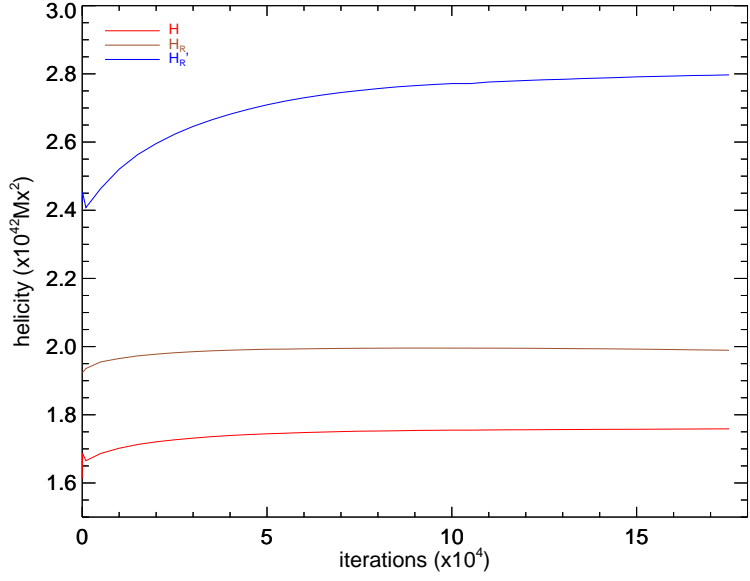}
\caption{Evolution of relative magnetic helicity as computed with the method of this article, $H$, using Equation~\ref{helca0}, and with the approximate ones, $H_R$ using Equation~\ref{helb1}, and $H_R'$ using Equation~\ref{helb2}.}
\label{helsavb}
\end{figure}

The second set of vector fields that we use in this approximate method is the ones from the exact method. We use again the same relation, namely
\begin{equation}
H_R'=\int_V\,\mathrm{d}V (\mathbfit{A}_{\rm DV}\cdot \mathbfit{B}_{\rm DV} - \mathbfit{A}_\mathrm{p,DV}\cdot \mathbfit{B}_\mathrm{p,DV})+\int_S\,\mathrm{d}S \chi_{\rm DV} B_{{\rm DV},r},
\label{helb2}
\end{equation}
with $\chi_{\rm DV}(\mathcal{P})=\int_\mathcal{O}^\mathcal{P} \mathrm{d}\mathbfit{l}\cdot(\mathbfit{A}_{\rm DV}-\mathbfit{A}_\mathrm{p,DV})$. The evolution of helicity obtained with these two approximate methods is plotted in Figure~\ref{helsavb}, along with the result of the exact method in the DVSt--DVSt gauge combination.

We note that the $H_R$ method exhibits a very similar pattern with the exact method. The helicity values are however on average $15\,\%$ higher than those of the exact method. This difference is due to the limitations of the formulation of Equation~\ref{helb} with respect to the finite-volume geometry used, as was pointed out in the Introduction. We see therefore that our helicity computation method improves, by the designated percent, with respect to the approximate method of Equation~\ref{helb}.

For the $H_R'$ method, we note that the general pattern of helicity evolution is similar to the other two methods. The obtained helicity values however, are (on average) higher by $55\,\%$ than the exact ones, and by $35\,\%$ than those of the $H_R$ method. This large difference is expected since the vector potentials of Equation~\ref{helb2} are in the DeVore gauge and they do not assume the condition $H_{\rm mix}=0$ that Equation~\ref{helb} implies, since $\mathbfit{A}_\mathrm{DV}\times\mathbfit{A}_\mathrm{p,DV}\neq0$ on the boundary. According also to the discussion in the Introduction, the relation of Equation~\ref{helb} depends on the gauges chosen for the vector potentials, unless the bounding surface is a flux surface. The use of Equation~\ref{helb} with vector potentials that do not respect its validity conditions can thus lead to differences in helicity of the order of $30-40\,\%$, at least in the case examined here.

We conclude that helicity calculations based on Equation~\ref{helb} are problematic when using finite volumes (in whatever coordinate system). The conditions of validity of Equation~\ref{helb} are rarely true in practice and so its use with finite-volume data should be avoided. Situations where it is safer to use this relation include, for example, the infinite-plane geometry originally used in \citet{BergerF84}.

\section{Discussion}
     \label{S-discussion} 

A method for computing relative magnetic helicity in spherical geometry was presented in this article. The necessity for such a method stems from the fact that magnetic helicity is an important quantity with many applications in the solar context, and that the natural coordinate system for the Sun is the spherical one. The developed method treats the generic problem of computing helicity in a spherical box, and it is thus superior to previous methods that either employed simplified conditions on the boundary for the magnetic fields and/or the vector potentials, or used approximations in the computations.

Testing of the developed method with semi-analytic NLFF field models showed that it is working very well, with comparable accuracy to corresponding calculations in the Cartesian case. More specifically, the potential magnetic field produced by the method is highly divergence-free, and also, the computed vector potentials reproduce the respective magnetic fields to a high degree.

Additionally, from the application on a data-driven NLFF field model of a solar AR we already see an improvement on the values of helicity compared to approximate methods used in the past. The specific approximate method that we examined here suffers from an important limitation. It assumes a certain choice for the gauge of the vector potentials, which is hard to enforce in finite volumes. As a result, the obtained helicity values with the different gauge choices can be quite different from each other, and also, from the method developed in this article.

In all cases where the magnetic field is in the spherical geometry and it is known in the whole volume, such as in MHD simulations or NLFF field reconstructions, the method presented here is expected to give the most accurate estimation of the structure's magnetic helicity, and additionally with minimal computational effort and resources.

\begin{acks}
E. Pariat and K. Moraitis acknowledge the support of the French Agence Nationale pour la Recherche through the HELISOL project, contract n$^\mathrm{o}$ ANR-15-CE31-0001. G. Valori acknowledges the support of the Leverhulme Trust Research Project Grant 2014-051. A. Savcheva was funded by NASA HSR grant NNX16AH87G.
\end{acks}

\section*{Disclosure of Potential Conflicts of Interest}

The authors declare that they have no conflicts of interest.

\bibliographystyle{spr-mp-sola} 
\bibliography{refs}

\begin{thebibliography}{37}
\ifx\bisbn     \undefined \def\bisbn  #1{ISBN #1}\fi
\ifx\binits    \undefined \def\binits#1{#1}\fi
\ifx\bauthor   \undefined \def\bauthor#1{#1}\fi
\ifx\batitle   \undefined \def\batitle#1{#1}\fi
\ifx\bjtitle   \undefined \def\bjtitle#1{\textit{#1}}\fi
\ifx\bvolume   \undefined \def\bvolume#1{\textbf{#1}}\fi
\ifx\byear     \undefined \def\byear#1{#1}\fi
\ifx\bissue    \undefined \def\bissue#1{#1}\fi
\ifx\bfpage    \undefined \def\bfpage#1{#1}\fi
\ifx\blpage    \undefined \def\blpage #1{#1}\fi
\ifx\burl      \undefined \def\burl#1{\textsf{#1}}\fi
\ifx\href      \undefined \def\href#1#2{\textsf{#2}}\fi
\ifx\betal     \undefined \def\betal{\textit{et al.}}\fi
\ifx\bctitle   \undefined \def\bctitle#1{#1}\fi
\ifx\beditor   \undefined \def\beditor#1{#1}\fi
\ifx\bbtitle   \undefined \def\bbtitle#1{\textit{#1}}\fi
\ifx\bedition  \undefined \def\bedition#1{#1}\fi
\ifx\bseriesno \undefined \def\bseriesno#1{\textbf{#1}}\fi
\ifx\blocation \undefined \def\blocation#1{#1}\fi
\ifx\bsertitle \undefined \def\bsertitle#1{\textit{#1}}\fi
\ifx\bsnm      \undefined \def\bsnm#1{#1}\fi
\ifx\bsuffix   \undefined \def\bsuffix#1{#1}\fi
\ifx\bparticle \undefined \def\bparticle#1{#1}\fi
\ifx\barticle  \undefined \def\barticle#1{}\fi
\ifx\binstitute  \undefined \def\binstitute#1{#1}\fi
\ifx\bpublisher  \undefined \def\bpublisher#1{#1}\fi
\ifx\doiurl    \undefined
  \def\doiurl#1{\href{http://dx.doi.org/#1}{\textsf{DOI}}}\fi
\ifx\arxivurl  \undefined
  \def\arxivurl#1{\href{http://arxiv.org/abs/#1}{\textsf{arXiv}}}\fi
\ifx\adsurl    \undefined
  \def\adsurl#1{\href{http://adsabs.harvard.edu/abs/#1}{\textsf{ADS}}}\fi
\ifx\botherref \undefined \def\botherref#1{}\fi
\ifx\url       \undefined \def\url#1{\textsf{#1}}\fi
\ifx\bchapter  \undefined \def\bchapter#1{}\fi
\ifx\bbook     \undefined \def\bbook#1{}\fi
\ifx\bcomment  \undefined \def\bcomment#1{#1}\fi
\ifx\oauthor   \undefined \def\oauthor#1{#1}\fi
\ifx\citeauthoryear \undefined\def \citeauthoryear#1{#1}\fi
\ifx\endbibitem\undefined \def\endbibitem{}\fi
\ifx\bconflocation  \undefined \def\bconflocation#1{#1} \fi

\bibitem[\protect\citeauthoryear{{Adams}}{1989}]{adams89}
\begin{barticle}
\bauthor{\bsnm{{Adams}}, \binits{J.}}:
\byear{1989},
\batitle{{MUDPACK: Multigrid Fortran Software for the Efficient Solution of
  Linear Elliptic Partial Differential Equations}}.
\bjtitle{Applied Math. Comput.}
\bvolume{34},
\bfpage{113}.
\doiurl{10.1016/0096-3003(89)90010-6}.
\end{barticle}
\endbibitem

\bibitem[\protect\citeauthoryear{{Amari} \textit{et~al.}}{2013}]{amari13}
\begin{barticle}
\bauthor{\bsnm{{Amari}}, \binits{T.}},
\bauthor{\bsnm{{Aly}}, \binits{J.-J.}},
\bauthor{\bsnm{{Canou}}, \binits{A.}},
\bauthor{\bsnm{{Mikic}}, \binits{Z.}}:
\byear{2013},
\batitle{{Reconstruction of the solar coronal magnetic field in spherical
  geometry}}.
\bjtitle{\aap}
\bvolume{553},
\bfpage{A43}.
\doiurl{10.1051/0004-6361/201220787}.
\adsurl{2013A\%26A...553A..43A}.
\end{barticle}
\endbibitem

\bibitem[\protect\citeauthoryear{{Berger} and {Field}}{1984}]{BergerF84}
\begin{barticle}
\bauthor{\bsnm{{Berger}}, \binits{M.A.}},
\bauthor{\bsnm{{Field}}, \binits{G.B.}}:
\byear{1984},
\batitle{{The topological properties of magnetic helicity}}.
\bjtitle{J. Fluid Mech.}
\bvolume{147},
\bfpage{133}.
\doiurl{10.1017/S0022112084002019}.
\adsurl{1984JFM...147..133B}.
\end{barticle}
\endbibitem

\bibitem[\protect\citeauthoryear{{Bobra}, {van Ballegooijen}, and
  {DeLuca}}{2008}]{bobra08}
\begin{barticle}
\bauthor{\bsnm{{Bobra}}, \binits{M.G.}},
\bauthor{\bsnm{{van Ballegooijen}}, \binits{A.A.}},
\bauthor{\bsnm{{DeLuca}}, \binits{E.E.}}:
\byear{2008},
\batitle{{Modeling Nonpotential Magnetic Fields in Solar Active Regions}}.
\bjtitle{\apj}
\bvolume{672},
\bfpage{1209}.
\doiurl{10.1086/523927}.
\adsurl{2008ApJ...672.1209B}.
\end{barticle}
\endbibitem

\bibitem[\protect\citeauthoryear{{Brandenburg} and
  {Subramanian}}{2005}]{brandenburg05}
\begin{barticle}
\bauthor{\bsnm{{Brandenburg}}, \binits{A.}},
\bauthor{\bsnm{{Subramanian}}, \binits{K.}}:
\byear{2005},
\batitle{{Astrophysical magnetic fields and nonlinear dynamo theory}}.
\bjtitle{Phys. Rep.}
\bvolume{417},
\bfpage{1}.
\doiurl{10.1016/j.physrep.2005.06.005}.
\adsurl{2005PhR...417....1B}.
\end{barticle}
\endbibitem

\bibitem[\protect\citeauthoryear{{Brown} \textit{et~al.}}{1999}]{canfield99}
\begin{bbook}
\bauthor{\bsnm{{Brown}}, \binits{M.}},
\bauthor{\bsnm{{Canfield}}, \binits{R.}},
\bauthor{\bsnm{{Field}}, \binits{G.}},
\bauthor{\bsnm{{Kulsrud}}, \binits{R.}},
\bauthor{\bsnm{{Pevtsov}}, \binits{A.}},
\bauthor{\bsnm{{Rosner}}, \binits{R.}},
\bauthor{\bsnm{{Seehafer}}, \binits{N.}}:
\byear{1999},
\bbtitle{{Magnetic helicity in space and laboratory plasmas: Editorial
  summary}}
\bseriesno{111},
\bpublisher{AGU},
\blocation{Washington},
\bfpage{301}.
\doiurl{10.1029/GM111p0301}.
\adsurl{1999GMS...111..301B}.
\end{bbook}
\endbibitem

\bibitem[\protect\citeauthoryear{{Dasgupta} \textit{et~al.}}{2002}]{dasgupta02}
\begin{barticle}
\bauthor{\bsnm{{Dasgupta}}, \binits{B.}},
\bauthor{\bsnm{{Janaki}}, \binits{M.S.}},
\bauthor{\bsnm{{Bhattacharyya}}, \binits{R.}},
\bauthor{\bsnm{{Dasgupta}}, \binits{P.}},
\bauthor{\bsnm{{Watanabe}}, \binits{T.}},
\bauthor{\bsnm{{Sato}}, \binits{T.}}:
\byear{2002},
\batitle{{Spheromak as a relaxed state with minimum dissipation}}.
\bjtitle{\pre}
\bvolume{65}(\bissue{4}),
\bfpage{046405}.
\doiurl{10.1103/PhysRevE.65.046405}.
\adsurl{2002PhRvE..65d6405D}.
\end{barticle}
\endbibitem

\bibitem[\protect\citeauthoryear{{DeVore}}{2000}]{devore00}
\begin{barticle}
\bauthor{\bsnm{{DeVore}}, \binits{C.R.}}:
\byear{2000},
\batitle{{Magnetic Helicity Generation by Solar Differential Rotation}}.
\bjtitle{\apj}
\bvolume{539},
\bfpage{944}.
\doiurl{10.1086/309274}.
\adsurl{2000ApJ...539..944D}.
\end{barticle}
\endbibitem

\bibitem[\protect\citeauthoryear{{Fan}}{2010}]{fan10}
\begin{barticle}
\bauthor{\bsnm{{Fan}}, \binits{Y.}}:
\byear{2010},
\batitle{{On the Eruption of Coronal Flux Ropes}}.
\bjtitle{\apj}
\bvolume{719},
\bfpage{728}.
\doiurl{10.1088/0004-637X/719/1/728}.
\adsurl{2010ApJ...719..728F}.
\end{barticle}
\endbibitem

\bibitem[\protect\citeauthoryear{{Fan}}{2016}]{fan16}
\begin{barticle}
\bauthor{\bsnm{{Fan}}, \binits{Y.}}:
\byear{2016},
\batitle{{Modeling the Initiation of the 2006 December 13 Coronal Mass Ejection
  in AR 10930: The Structure and Dynamics of the Erupting Flux Rope}}.
\bjtitle{\apj}
\bvolume{824},
\bfpage{93}.
\doiurl{10.3847/0004-637X/824/2/93}.
\adsurl{2016ApJ...824...93F}.
\end{barticle}
\endbibitem

\bibitem[\protect\citeauthoryear{{Finn} and {Antonsen}}{1985}]{fa85}
\begin{barticle}
\bauthor{\bsnm{{Finn}}, \binits{J.M.}},
\bauthor{\bsnm{{Antonsen}}, \binits{T.M.}}:
\byear{1985},
\batitle{{Magnetic helicity: What is it and what is it good for?}}
\bjtitle{Comments Plasma Phys. Control. Fusion}
\bvolume{9},
\bfpage{111}.
\end{barticle}
\endbibitem

\bibitem[\protect\citeauthoryear{{Gilchrist} and
  {Wheatland}}{2014}]{Gilchrist2014}
\begin{barticle}
\bauthor{\bsnm{{Gilchrist}}, \binits{S.A.}},
\bauthor{\bsnm{{Wheatland}}, \binits{M.S.}}:
\byear{2014},
\batitle{{Nonlinear Force-Free Modeling of the Corona in Spherical
  Coordinates}}.
\bjtitle{\solphys}
\bvolume{289},
\bfpage{1153}.
\doiurl{10.1007/s11207-013-0406-5}.
\adsurl{2014SoPh..289.1153G}.
\end{barticle}
\endbibitem

\bibitem[\protect\citeauthoryear{{Ji}, {Prager}, and {Sarff}}{1995}]{ji95}
\begin{barticle}
\bauthor{\bsnm{{Ji}}, \binits{H.}},
\bauthor{\bsnm{{Prager}}, \binits{S.C.}},
\bauthor{\bsnm{{Sarff}}, \binits{J.S.}}:
\byear{1995},
\batitle{{Conservation of Magnetic Helicity during Plasma Relaxation}}.
\bjtitle{Phys. Rev. Lett.}
\bvolume{74},
\bfpage{2945}.
\doiurl{10.1103/PhysRevLett.74.2945}.
\adsurl{1995PhRvL..74.2945J}.
\end{barticle}
\endbibitem

\bibitem[\protect\citeauthoryear{{Karpen} \textit{et~al.}}{2017}]{karpen17}
\begin{barticle}
\bauthor{\bsnm{{Karpen}}, \binits{J.T.}},
\bauthor{\bsnm{{DeVore}}, \binits{C.R.}},
\bauthor{\bsnm{{Antiochos}}, \binits{S.K.}},
\bauthor{\bsnm{{Pariat}}, \binits{E.}}:
\byear{2017},
\batitle{{Reconnection-Driven Coronal-Hole Jets with Gravity and Solar Wind}}.
\bjtitle{\apj}
\bvolume{834},
\bfpage{62}.
\doiurl{10.3847/1538-4357/834/1/62}.
\adsurl{2017ApJ...834...62K}.
\end{barticle}
\endbibitem

\bibitem[\protect\citeauthoryear{{Kliem} \textit{et~al.}}{2013}]{kliem13}
\begin{barticle}
\bauthor{\bsnm{{Kliem}}, \binits{B.}},
\bauthor{\bsnm{{Su}}, \binits{Y.N.}},
\bauthor{\bsnm{{van Ballegooijen}}, \binits{A.A.}},
\bauthor{\bsnm{{DeLuca}}, \binits{E.E.}}:
\byear{2013},
\batitle{{Magnetohydrodynamic Modeling of the Solar Eruption on 2010 April 8}}.
\bjtitle{\apj}
\bvolume{779},
\bfpage{129}.
\doiurl{10.1088/0004-637X/779/2/129}.
\adsurl{2013ApJ...779..129K}.
\end{barticle}
\endbibitem

\bibitem[\protect\citeauthoryear{{Low} and {Lou}}{1990}]{lowlou90}
\begin{barticle}
\bauthor{\bsnm{{Low}}, \binits{B.C.}},
\bauthor{\bsnm{{Lou}}, \binits{Y.Q.}}:
\byear{1990},
\batitle{{Modeling solar force-free magnetic fields}}.
\bjtitle{\apj}
\bvolume{352},
\bfpage{343}.
\doiurl{10.1086/168541}.
\adsurl{1990ApJ...352..343L}.
\end{barticle}
\endbibitem

\bibitem[\protect\citeauthoryear{{Masson}, {Antiochos}, and
  {DeVore}}{2013}]{masson13}
\begin{barticle}
\bauthor{\bsnm{{Masson}}, \binits{S.}},
\bauthor{\bsnm{{Antiochos}}, \binits{S.K.}},
\bauthor{\bsnm{{DeVore}}, \binits{C.R.}}:
\byear{2013},
\batitle{{A Model for the Escape of Solar-flare-accelerated Particles}}.
\bjtitle{\apj}
\bvolume{771},
\bfpage{82}.
\doiurl{10.1088/0004-637X/771/2/82}.
\adsurl{2013ApJ...771...82M}.
\end{barticle}
\endbibitem

\bibitem[\protect\citeauthoryear{{Moraitis} \textit{et~al.}}{2014}]{moraitis14}
\begin{barticle}
\bauthor{\bsnm{{Moraitis}}, \binits{K.}},
\bauthor{\bsnm{{Tziotziou}}, \binits{K.}},
\bauthor{\bsnm{{Georgoulis}}, \binits{M.K.}},
\bauthor{\bsnm{{Archontis}}, \binits{V.}}:
\byear{2014},
\batitle{{Validation and Benchmarking of a Practical Free Magnetic Energy and
  Relative Magnetic Helicity Budget Calculation in Solar Magnetic Structures}}.
\bjtitle{\solphys}
\bvolume{289},
\bfpage{4453}.
\doiurl{10.1007/s11207-014-0590-y}.
\adsurl{2014SoPh..289.4453M}.
\end{barticle}
\endbibitem

\bibitem[\protect\citeauthoryear{{Pariat} \textit{et~al.}}{2015}]{pariat15}
\begin{barticle}
\bauthor{\bsnm{{Pariat}}, \binits{E.}},
\bauthor{\bsnm{{Valori}}, \binits{G.}},
\bauthor{\bsnm{{D{\'e}moulin}}, \binits{P.}},
\bauthor{\bsnm{{Dalmasse}}, \binits{K.}}:
\byear{2015},
\batitle{{Testing magnetic helicity conservation in a solar-like active
  event}}.
\bjtitle{\aap}
\bvolume{580},
\bfpage{A128}.
\doiurl{10.1051/0004-6361/201525811}.
\adsurl{2015A\%26A...580A.128P}.
\end{barticle}
\endbibitem

\bibitem[\protect\citeauthoryear{{Pariat} \textit{et~al.}}{2017}]{pariat17}
\begin{barticle}
\bauthor{\bsnm{{Pariat}}, \binits{E.}},
\bauthor{\bsnm{{Leake}}, \binits{J.E.}},
\bauthor{\bsnm{{Valori}}, \binits{G.}},
\bauthor{\bsnm{{Linton}}, \binits{M.G.}},
\bauthor{\bsnm{{Zuccarello}}, \binits{F.P.}},
\bauthor{\bsnm{{Dalmasse}}, \binits{K.}}:
\byear{2017},
\batitle{{Relative magnetic helicity as a diagnostic of solar eruptivity}}.
\bjtitle{\aap}
\bvolume{601},
\bfpage{A125}.
\doiurl{10.1051/0004-6361/201630043}.
\adsurl{2017A\%26A...601A.125P}.
\end{barticle}
\endbibitem

\bibitem[\protect\citeauthoryear{{Polito} \textit{et~al.}}{2017}]{polito17}
\begin{barticle}
\bauthor{\bsnm{{Polito}}, \binits{V.}},
\bauthor{\bsnm{{Del Zanna}}, \binits{G.}},
\bauthor{\bsnm{{Valori}}, \binits{G.}},
\bauthor{\bsnm{{Pariat}}, \binits{E.}},
\bauthor{\bsnm{{Mason}}, \binits{H.E.}},
\bauthor{\bsnm{{Dud{\'{\i}}k}}, \binits{J.}},
\bauthor{\bsnm{{Janvier}}, \binits{M.}}:
\byear{2017},
\batitle{{Analysis and modelling of recurrent solar flares observed with
  Hinode/EIS on March 9, 2012}}.
\bjtitle{\aap}
\bvolume{601},
\bfpage{A39}.
\doiurl{10.1051/0004-6361/201629703}.
\adsurl{2017A\%26A...601A..39P}.
\end{barticle}
\endbibitem

\bibitem[\protect\citeauthoryear{{Rust}}{1994}]{rust94}
\begin{barticle}
\bauthor{\bsnm{{Rust}}, \binits{D.M.}}:
\byear{1994},
\batitle{{Spawning and shedding helical magnetic fields in the solar
  atmosphere}}.
\bjtitle{\grl}
\bvolume{21},
\bfpage{241}.
\doiurl{10.1029/94GL00003}.
\adsurl{1994GeoRL..21..241R}.
\end{barticle}
\endbibitem

\bibitem[\protect\citeauthoryear{{Savcheva} \textit{et~al.}}{2015}]{savcheva15}
\begin{barticle}
\bauthor{\bsnm{{Savcheva}}, \binits{A.}},
\bauthor{\bsnm{{Pariat}}, \binits{E.}},
\bauthor{\bsnm{{McKillop}}, \binits{S.}},
\bauthor{\bsnm{{McCauley}}, \binits{P.}},
\bauthor{\bsnm{{Hanson}}, \binits{E.}},
\bauthor{\bsnm{{Su}}, \binits{Y.}},
\bauthor{\bsnm{{Werner}}, \binits{E.}},
\bauthor{\bsnm{{DeLuca}}, \binits{E.E.}}:
\byear{2015},
\batitle{{The Relation between Solar Eruption Topologies and Observed Flare
  Features. I. Flare Ribbons}}.
\bjtitle{\apj}
\bvolume{810},
\bfpage{96}.
\doiurl{10.1088/0004-637X/810/2/96}.
\adsurl{2015ApJ...810...96S}.
\end{barticle}
\endbibitem

\bibitem[\protect\citeauthoryear{{Savcheva} \textit{et~al.}}{2016}]{savcheva16}
\begin{barticle}
\bauthor{\bsnm{{Savcheva}}, \binits{A.}},
\bauthor{\bsnm{{Pariat}}, \binits{E.}},
\bauthor{\bsnm{{McKillop}}, \binits{S.}},
\bauthor{\bsnm{{McCauley}}, \binits{P.}},
\bauthor{\bsnm{{Hanson}}, \binits{E.}},
\bauthor{\bsnm{{Su}}, \binits{Y.}},
\bauthor{\bsnm{{DeLuca}}, \binits{E.E.}}:
\byear{2016},
\batitle{{The Relation between Solar Eruption Topologies and Observed Flare
  Features. II. Dynamical Evolution}}.
\bjtitle{\apj}
\bvolume{817},
\bfpage{43}.
\doiurl{10.3847/0004-637X/817/1/43}.
\adsurl{2016ApJ...817...43S}.
\end{barticle}
\endbibitem

\bibitem[\protect\citeauthoryear{{Schrijver}
  \textit{et~al.}}{2006}]{schrijver06}
\begin{barticle}
\bauthor{\bsnm{{Schrijver}}, \binits{C.J.}},
\bauthor{\bsnm{{De Rosa}}, \binits{M.L.}},
\bauthor{\bsnm{{Metcalf}}, \binits{T.R.}},
\bauthor{\bsnm{{Liu}}, \binits{Y.}},
\bauthor{\bsnm{{McTiernan}}, \binits{J.}},
\bauthor{\bsnm{{R{\'e}gnier}}, \binits{S.}},
\bauthor{\bsnm{{Valori}}, \binits{G.}},
\bauthor{\bsnm{{Wheatland}}, \binits{M.S.}},
\bauthor{\bsnm{{Wiegelmann}}, \binits{T.}}:
\byear{2006},
\batitle{{Nonlinear Force-Free Modeling of Coronal Magnetic Fields Part I: A
  Quantitative Comparison of Methods}}.
\bjtitle{\solphys}
\bvolume{235},
\bfpage{161}.
\doiurl{10.1007/s11207-006-0068-7}.
\adsurl{2006SoPh..235..161S}.
\end{barticle}
\endbibitem

\bibitem[\protect\citeauthoryear{Swarztrauber and Sweet}{1979}]{ss79}
\begin{barticle}
\bauthor{\bsnm{Swarztrauber}, \binits{P.N.}},
\bauthor{\bsnm{Sweet}, \binits{R.A.}}:
\byear{1979},
\batitle{Algorithm 541: Efficient fortran subprograms for the solution of
  separable elliptic partial differential equations}.
\bjtitle{ACM Trans. Math. Softw.}
\bvolume{5},
\bfpage{352}.
\doiurl{10.1145/355841.355850}.
\end{barticle}
\endbibitem

\bibitem[\protect\citeauthoryear{{Taylor}}{1974}]{taylor74}
\begin{barticle}
\bauthor{\bsnm{{Taylor}}, \binits{J.B.}}:
\byear{1974},
\batitle{{Relaxation of Toroidal Plasma and Generation of Reverse Magnetic
  Fields}}.
\bjtitle{Phys. Rev. Lett.}
\bvolume{33},
\bfpage{1139}.
\doiurl{10.1103/PhysRevLett.33.1139}.
\adsurl{1974PhRvL..33.1139T}.
\end{barticle}
\endbibitem

\bibitem[\protect\citeauthoryear{{Thalmann}, {Inhester}, and
  {Wiegelmann}}{2011}]{thalmann11}
\begin{barticle}
\bauthor{\bsnm{{Thalmann}}, \binits{J.K.}},
\bauthor{\bsnm{{Inhester}}, \binits{B.}},
\bauthor{\bsnm{{Wiegelmann}}, \binits{T.}}:
\byear{2011},
\batitle{{Estimating the Relative Helicity of Coronal Magnetic Fields}}.
\bjtitle{\solphys}
\bvolume{272},
\bfpage{243}.
\doiurl{10.1007/s11207-011-9826-2}.
\adsurl{2011SoPh..272..243T}.
\end{barticle}
\endbibitem

\bibitem[\protect\citeauthoryear{{Tziotziou}
  \textit{et~al.}}{2014}]{tziotziou14}
\begin{barticle}
\bauthor{\bsnm{{Tziotziou}}, \binits{K.}},
\bauthor{\bsnm{{Moraitis}}, \binits{K.}},
\bauthor{\bsnm{{Georgoulis}}, \binits{M.K.}},
\bauthor{\bsnm{{Archontis}}, \binits{V.}}:
\byear{2014},
\batitle{{Validation of the magnetic energy vs. helicity scaling in solar
  magnetic structures}}.
\bjtitle{\aapl}
\bvolume{570},
\bfpage{L1}.
\doiurl{10.1051/0004-6361/201424864}.
\adsurl{2014A\%26A...570L...1T}.
\end{barticle}
\endbibitem

\bibitem[\protect\citeauthoryear{{Valori}, {D{\'e}moulin}, and
  {Pariat}}{2012}]{val12}
\begin{barticle}
\bauthor{\bsnm{{Valori}}, \binits{G.}},
\bauthor{\bsnm{{D{\'e}moulin}}, \binits{P.}},
\bauthor{\bsnm{{Pariat}}, \binits{E.}}:
\byear{2012},
\batitle{{Comparing Values of the Relative Magnetic Helicity in Finite
  Volumes}}.
\bjtitle{\solphys}
\bvolume{278},
\bfpage{347}.
\doiurl{10.1007/s11207-012-0044-3}.
\adsurl{2012SoPh..278..347V}.
\end{barticle}
\endbibitem

\bibitem[\protect\citeauthoryear{{Valori} \textit{et~al.}}{2013}]{val13}
\begin{barticle}
\bauthor{\bsnm{{Valori}}, \binits{G.}},
\bauthor{\bsnm{{D{\'e}moulin}}, \binits{P.}},
\bauthor{\bsnm{{Pariat}}, \binits{E.}},
\bauthor{\bsnm{{Masson}}, \binits{S.}}:
\byear{2013},
\batitle{{Accuracy of magnetic energy computations}}.
\bjtitle{\aap}
\bvolume{553},
\bfpage{A38}.
\doiurl{10.1051/0004-6361/201220982}.
\adsurl{2013A\%26A...553A..38V}.
\end{barticle}
\endbibitem

\bibitem[\protect\citeauthoryear{{Valori} \textit{et~al.}}{2016}]{valori16}
\begin{barticle}
\bauthor{\bsnm{{Valori}}, \binits{G.}},
\bauthor{\bsnm{{Pariat}}, \binits{E.}},
\bauthor{\bsnm{{Anfinogentov}}, \binits{S.}},
\bauthor{\bsnm{{Chen}}, \binits{F.}},
\bauthor{\bsnm{{Georgoulis}}, \binits{M.K.}},
\bauthor{\bsnm{{Guo}}, \binits{Y.}},
\bauthor{\bsnm{{Liu}}, \binits{Y.}},
\bauthor{\bsnm{{Moraitis}}, \binits{K.}},
\bauthor{\bsnm{{Thalmann}}, \binits{J.K.}},
\bauthor{\bsnm{{Yang}}, \binits{S.}}:
\byear{2016},
\batitle{{Magnetic Helicity Estimations in Models and Observations of the Solar
  Magnetic Field. Part I: Finite Volume Methods}}.
\bjtitle{\ssr}
\bvolume{201},
\bfpage{147}.
\doiurl{10.1007/s11214-016-0299-3}.
\adsurl{2016SSRv..201..147V}.
\end{barticle}
\endbibitem

\bibitem[\protect\citeauthoryear{{van Ballegooijen}}{2004}]{vanB04}
\begin{barticle}
\bauthor{\bsnm{{van Ballegooijen}}, \binits{A.A.}}:
\byear{2004},
\batitle{{Observations and Modeling of a Filament on the Sun}}.
\bjtitle{\apj}
\bvolume{612},
\bfpage{519}.
\doiurl{10.1086/422512}.
\adsurl{2004ApJ...612..519V}.
\end{barticle}
\endbibitem

\bibitem[\protect\citeauthoryear{{Wheatland}, {Sturrock}, and
  {Roumeliotis}}{2000}]{wheatland00}
\begin{barticle}
\bauthor{\bsnm{{Wheatland}}, \binits{M.S.}},
\bauthor{\bsnm{{Sturrock}}, \binits{P.A.}},
\bauthor{\bsnm{{Roumeliotis}}, \binits{G.}}:
\byear{2000},
\batitle{{An Optimization Approach to Reconstructing Force-free Fields}}.
\bjtitle{\apj}
\bvolume{540},
\bfpage{1150}.
\doiurl{10.1086/309355}.
\adsurl{2000ApJ...540.1150W}.
\end{barticle}
\endbibitem

\bibitem[\protect\citeauthoryear{{Woltjer}}{1958}]{woltjer58}
\begin{barticle}
\bauthor{\bsnm{{Woltjer}}, \binits{L.}}:
\byear{1958},
\batitle{{A Theorem on Force-Free Magnetic Fields}}.
\bjtitle{Proc. Nat. Acad. Science}
\bvolume{44},
\bfpage{489}.
\doiurl{10.1073/pnas.44.6.489}.
\adsurl{1958PNAS...44..489W}.
\end{barticle}
\endbibitem

\bibitem[\protect\citeauthoryear{{Yang} \textit{et~al.}}{2013}]{yang13}
\begin{barticle}
\bauthor{\bsnm{{Yang}}, \binits{S.}},
\bauthor{\bsnm{{B{\"u}chner}}, \binits{J.}},
\bauthor{\bsnm{{Santos}}, \binits{J.C.}},
\bauthor{\bsnm{{Zhang}}, \binits{H.}}:
\byear{2013},
\batitle{{Evolution of Relative Magnetic Helicity: Method of Computation and
  Its Application to a Simulated Solar Corona above an Active Region}}.
\bjtitle{\solphys}
\bvolume{283},
\bfpage{369}.
\doiurl{10.1007/s11207-013-0236-5}.
\adsurl{2013SoPh..283..369Y}.
\end{barticle}
\endbibitem

\bibitem[\protect\citeauthoryear{{Yeates} and {Hornig}}{2016}]{yeates16}
\begin{barticle}
\bauthor{\bsnm{{Yeates}}, \binits{A.R.}},
\bauthor{\bsnm{{Hornig}}, \binits{G.}}:
\byear{2016},
\batitle{{The global distribution of magnetic helicity in the solar corona}}.
\bjtitle{\aap}
\bvolume{594},
\bfpage{A98}.
\doiurl{10.1051/0004-6361/201629122}.
\adsurl{2016A\%26A...594A..98Y}.
\end{barticle}
\endbibitem

\end{thebibliography}

\end{article} 

\end{document}